\hfuzz=3pt
\hoffset=-.15in 
\def\page{\footline={\ifnum\pageno=1 \hfil
\else\hss\twelverm\folio\hss\fi}}
\def\unpage{\footline={\ifnum\pageno=0 \hfil
\else\hss\twelverm\folio\hss\fi}}

\def\mo#1{\kern-5.5pt #1}
\def\beginpreprint{
\abovedisplayskip=12pt plus6pt minus5pt
\belowdisplayskip=12pt plus6pt minus5pt
\page\twelvepoint\rm
\raggedbottom
\begintitlepage
\def\endtitlepage{\eject}
\def\keywords##1{\parindent=1in{\item{\it Keywords:} ##1}\parindent=20pt}
\def\section##1\par{\vskip.4in \centerline{\bf ##1}\par\bigskip}
\def\sectiona##1\par{\vskip.4in \centerline{\bf ##1}\par}
\def\sectionb##1\par{\centerline{\bf ##1}\par\bigskip}
\def\subsection##1\par{\bigbreak \centerline{\it
##1}\par\medskip}
\def\subsectiona##1\par{\bigbreak\centerline{\it ##1}}
\def\subsectionb##1\par{\centerline{\it ##1}\par\medskip}
\def\subsubsection##1\par{\bigbreak \noindent{\it ##1}\par}
\def\subsubsectiona##1\par{\bigbreak\noindent{\it ##1}}
\def\subsubsectionb##1\par{\noindent{\it ##1}\par\smallskip}
\def\acknow{\vskip.4in plus4pt minus4pt\centerline{\bf
Acknowledgements}\smallskip}
\def\refs{\vskip.4in plus4pt minus4pt \centerline{\bf REFERENCES}\par
\bigskip\beginrefs}
\def\references{\vskip.4in plus4pt minus 4pt \centerline{\bf REFERENCES}\par
\bigskip\beginrefs}
\def\refline{\underbar{\hskip.75in}.\ }
\def\figcaps{\beginfigcaps}}

\def\begintitlepage{\doublespace\null\vfil
\def\title##1\par{\centerline{\bf##1}\par}
\def\author##1\par{\vskip.4in{\centerline{\it ##1}}\par}
\def\affil##1\par{\centerline{##1}\par}
\def\tobe##1##2\par{\centerline{To be published in the
{##1} issue of the {\it##2}}}
\def\recacc##1;##2\par{\centerline{Received \underbar{~~##1~~}; Accepted
\underbar{~~##2~~}}}
\def\gap{\hskip1in}
\def\esa{Affiliated to the Astrophysics Division,
Space Science Department, ESA.}
}
\long\def\abstract#1{\tenpoint\rm\singlespace
\output{\onepageout{\unvbox255}}
  {\centerline{\bf ABSTRACT}}\smallskip
  {\narrower\narrower{#1}\par}
   \begindoublecolumns
   \baselineskip=12pt plus.8pt minus.8pt
   \topskip=12pt
   \parskip=0pt plus.8pt minus.8pt
   \hyphenpenalty=-50\pretolerance=50\tolerance=600
   \doublehyphendemerits=750\finalhyphendemerits=-100000
   \brokenpenalty=-100
   \def\footline{\vskip.3in\hbox to 6.5in{\hss\tenrm\folio\hss}}
    \pageno=2
   }
\vfuzz=65pt
\catcode`@=11
\newdimen\pagewidth \newdimen\pageheight \newdimen\ruleht
\hsize=6.5in \vsize=9in \maxdepth=2.2pt
\pagewidth=\hsize \pageheight=\vsize \ruleht=.5pt
\def\footnote#1{\edef\@sf{\spacefactor\the\spacefactor}#1\@sf
  \insert\footins\bgroup\tenpoint
  \interlinepenalty100 \let\par=\endgraf
  \leftskip=0pt \rightskip=0pt
  \splittopskip=12pt plus 1pt minus 1pt \floatingpenalty=20000
  \smallskip\item{#1}\bgroup\strut\aftergroup\@foot\let\next}
\skip\footins=14pt plus 2pt minus 4pt 
\dimen\footins=30pc 
\def\onepageout#1{\shipout\vbox{
\offinterlineskip
\vbox to \pageheight{
#1
\ifvoid\footins\else
\vskip\skip\footins \kern-3pt
\hrule height\ruleht width3.15in \kern-\ruleht \kern3pt
\unvbox\footins\fi
\boxmaxdepth=\maxdepth}
\vbox to -3pc{\footline}
}
\advancepageno}
\newbox\partialpage
\def\begindoublecolumns{\begingroup
  \output={\global\setbox\partialpage=\vbox{\unvbox255\bigskip}}\eject
  \output={\doublecolumnout} \hsize=3.15in \vsize=18in}
\def\enddoublecolumns{\output={\balancecolumns}\eject
  \endgroup \pagegoal=\vsize}
\def\doublecolumnout{\splittopskip=\topskip \splitmaxdepth=\maxdepth
  \dimen@=9in \advance\dimen@ by-\ht\partialpage
  \setbox0=\vsplit255 to\dimen@ \setbox2=\vsplit255 to\dimen@
  \onepageout\pagesofar \unvbox255 \penalty\outputpenalty}
\def\pagesofar{\unvbox\partialpage
  \wd0=\hsize \wd2=\hsize \hbox to\pagewidth{\box0\hfil\box2}}
\def\balancecolumns{\setbox0=\vbox{\unvbox255} \dimen@=\ht0
  \advance\dimen@ by\topskip \advance\dimen@ by-\baselineskip
  \divide\dimen@ by2 \splittopskip=\topskip
  {\vbadness=10000 \loop \global\setbox3=\copy0
   \global\setbox1=\vsplit3 to\dimen@
   \ifdim\ht3>\dimen@ \global\advance\dimen@ by1pt \repeat}
\setbox0=\vbox to\dimen@{\unvbox1}
\setbox2=\vbox to\dimen@{
\dimen2=\dp3 \unvbox3 \kern-\dimen2 \vfil}
\pagesofar}
\def\ahead#1\smallskip{{\vfil\eject{\centerline{\bf#1}}
\smallskip}{\message#1}}
\def\bhead#1\smallskip{{\bigbreak{\centerline{\it#1}}\smallskip}
{\message#1}}
\def\chead#1\par{{\bigbreak\noindent{\it#1}\par} {\message#1}}
\def\head#1. #2\par{\medbreak\centerline{{\bf#1.\enspace}{\it#2}}
\par\medbreak}
\def\levelone#1~ ~ #2\smallskip{\noindent#1~ ~ {\bf#2}\smallskip}
\def\leveltwo#1~ ~ #2\smallskip{\noindent#1~ ~ {\it#2}\smallskip}
\def\levelthree#1~ ~ #2\smallskip{\noindent#1~ ~ {#2}\smallskip}

\def\m{^m\kern-7pt .\kern+3.5pt}
\def\p{^{\prime\prime}\kern-2.1mm .\kern+.6mm}
\def\pone{\kern.1mm^{\prime}\kern-1.2mm .\kern+.3mm}
\def\dpoint{^d\kern-1.5mm .\kern+.3mm}
\def\cpoint{^{\circ}\kern-1.7mm.\kern+.35mm}
\def\hpoint{^h\kern-2.1mm .\kern+.6mm}
\def\y{^y\kern-1.9mm .\kern+.3mm}
\def\s{^s\kern-1.7mm .\kern+.3mm}

\def\apgt{\ {\raise-.5ex\hbox{$\buildrel>\over\sim$}}\ }
\def\aplt{\ {\raise-.5ex\hbox{$\buildrel<\over\sim$}}\ }
\def\deg{^{\circ}}

\def\hup{^{h}\kern-2.1mm .\kern+.6mm}

\def\ie{{\it i.e.,}\ }
\def\eg{{\it e.g.,}\ }

\def\sqr#1#2{{\vcenter{\hrule height.#2pt
\hbox{\vrule width.#2pt height#1pt \kern#1pt
\vrule width.#2pt}
\hrule height.#2pt}}}

\def\trule{\vskip6pt\hrule\vskip2pt\hrule\vskip6pt}
\def\mrule{\noalign{\vskip6pt\hrule\vskip6pt}}
\def\brule{\noalign{\vskip6pt\hrule}}

\def\etal{{\it et~al.\ }}

\def\beginfigcaps{\singlespace
  \def\footline{\vskip.3in\hbox to 6.5in{\hss\tenrm\folio\hss}}
\vfil\eject
\begingroup
\parindent=0pt\frenchspacing
\parskip=1pt plus 1pt minus 1pt
\def\fig##1---##2\par{\vfill\hangindent=1.6pc\bf ##1\rm ---##2\eject}
\def\footline{}
}
\def\endfigcaps{\endgroup}
\def\beginrefs{\begingroup\parindent=0pt\frenchspacing
\parskip=1pt plus 1pt minus 1pt\interlinepenalty=1000
\everypar={\hangindent=1.6pc}}
\def\endrefs{\enddoublecolumns\endgroup
}
\def\misref#1,#2,{{\it #1}, {\bf#2},}
\def\ajmisref#1,#2{{\it #1}, {\bf#2},}
\def\nature#1,{{\it Nature}, {\bf#1},}
\def\ajnature#1,{{\it Nature}, {\bf#1},}
\def\aa#1,{{\it Astr.\ Ap.,\ }{\bf#1},}

\def\ajaa#1,{{\it Astron.\ Astrophys.,\ }{\bf#1},}

\def\aalet#1,{{\it Astr.\ Ap.\ (Letters),\ }{\bf#1},}

\def\ajaalet#1,{{\it Astron. Astrophys. (Letters),\ }{\bf#1},}

\def\aasup#1,{{\it Astr. Ap. Suppl.,\ }{\bf#1},}

\def\ajaasup#1,{{\it Astron. Astrophys. Suppl.,\ }{\bf#1},}

\def\aass#1,{{\it Astr. Ap. Suppl. Ser.,\ }{\bf#1},}

\def\aj#1,{{\it A.~J.,\ }{\bf#1},}

\def\ajaj#1,{{\it Astron.~J.,\ }{\bf#1},}

\def\apj#1,{{\it Ap.~J.,\  }{\bf#1},}

\def\ajapj#1,{{\it Astrophys.~J.,\ }{\bf#1},}

\def\apjlet#1,{{\it Ap.~J. (Letters),\ }{\bf#1},}

\def\ajapjlet#1,{{\it Astrophys. J. Lett.,\ }{\bf#1},}

\def\apjsup#1,{{\it Ap.~J.~Suppl.,\ }{\bf#1},}

\def\ajapjsup#1,{{\it Astrophys. J. Suppl.,\ }{\bf#1},}

\def\araa#1,{{\it Ann.\ Rev.\ A.~A.,\ }{\bf#1},}

\def\ajaraa#1,{{\it Ann.\ Rev.\ A.~A.,\ }{\bf#1},}

\def\baas#1,{{\it B.A.A.S.,\ }{\bf#1},}

\def\ajbaas#1,{{\it B.A.A.S.,\ }{\bf#1},}

\def\mnras#1,{{\it M.N.R.A.S.,\ }{\bf#1},}

\def\ajmnras#1,{{\it M.N.R.A.S.,\ }{\bf#1},}

\def\pasp#1,{{\it Pub.~A.S.P.,\ }{\bf#1},}

\def\ajpasp#1,{{\it P.A.S.P.,\ }{\bf#1},}

\font\sevenrm=cmr7
\font\eightrm=cmr8

\font\tenrm=cmr10

\font\twelverm=cmr12
\font\eighteenrm=cmr10 scaled \magstep3
\def\twelvepoint{
\font\twelverm=cmr12
\font\twelvei=cmmi12
\font\twelvesy=cmsy10 scaled\magstep1
\font\tenrm=cmr10
\font\teni=cmmi10
\font\tensy=cmsy10
\font\sevenrm=cmr7
\font\seveni=cmmi7
\font\sevensy=cmsy7
\font\it=cmti12
\font\bf=cmbx12 
\font\bi=cmbxsl10 scaled \magstep1
\textfont0= \twelverm \scriptfont0=\tenrm
\scriptscriptfont0=\sevenrm
\def\rm{\fam0 \twelverm}
\textfont1=\twelvei  \scriptfont1=\teni
\scriptscriptfont1=\seveni
\def\mit{\fam1 } \def\oldstyle{\fam1 \twelvei}
\textfont2=\twelvesy \scriptfont2=\tensy
\scriptscriptfont2=\sevensy
\def\doublespace{\baselineskip=24pt\lineskip=0pt
\lineskiplimit=-5pt}
\def\singlespace{\baselineskip=13.5pt\lineskip=0pt
\lineskiplimit=-5pt}
\def\oneandahalf{\baselineskip=18pt\lineskip=0pt
\lineskiplimit=-5pt}
}

\def\tenpoint{
\font\tenrm=cmr10
\font\teni=cmmi10
\font\ten=cmsy10
\font\teni=cmmi10
\font\sevensy=cmsy7
\font\sevenrm=cmr7
\font\seveni=cmmi7
\font\fivei=cmmi5
\font\fivesy=cmsy5
\font\it=cmti10
\font\bf=cmb10
\font\bi=cmbxti10
\font\sl=cmsl10
\textfont0= \tenrm \scriptfont0=\sevenrm
\scriptscriptfont0=\fiverm
\def\rm{\fam0 \tenrm}
\textfont1=\teni  \scriptfont1=\seveni
\scriptscriptfont1=\fivei
\def\mit{\fam1 } \def\oldstyle{\fam1 \teni}
\textfont2=\tensy \scriptfont2=\sevensy
\scriptscriptfont2=\fivesy
\def\doublespace{\baselineskip=24pt\lineskip=0pt
\lineskiplimit=-5pt}
\def\singlespace{\baselineskip=12pt\lineskip=0pt
\lineskiplimit=-5pt}
\def\oneandahalf{\baselineskip=18pt\lineskip=0pt
\lineskiplimit=-5pt}
}

\def\eightpoint{
\font\eightrm=cmr8
\font\eighti=cmmi8
\font\eightsy=cmsy8
\font\sixi=cmmi6
\font\sixsy=cmsy6
\font\sixrm=cmr6
\font\fivei=cmmi5
\font\fivesy=cmsy5
\font\it=cmti8
\font\bf=cmbx8
\font\sl=cmsl8
\textfont0= \eightrm \scriptfont0=\sixrm
\scriptscriptfont0=\fiverm
\def\rm{\fam0 \eightrm}
\textfont1=\eighti  \scriptfont1=\sixi
\scriptscriptfont1=\fivei
\def\mit{\fam1 } \def\oldstyle{\fam1 \eighti}
\textfont2=\eightsy \scriptfont2=\sixsy
\scriptscriptfont2=\fivesy
\def\doublespace{\baselineskip=18pt\lineskip=0pt
\lineskiplimit=-5pt}
\def\singlespace{\baselineskip=9pt\lineskip=0pt
\lineskiplimit=-5pt}
\def\oneandahalf{\baselineskip=15pt\lineskip=0pt
\lineskiplimit=-5pt}
\def\p{^{\prime\prime}\kern-1.5mm .\kern+.3mm}
}
\def\eighteenpoint{
\font\eighteenrm=cmr10 scaled\magstep3
\font\eighteeni=cmmi10 scaled\magstep3
\font\eighteensy=cmsy10 scaled\magstep3
\font\eighteenrm=cmr10 scaled\magstep3
\font\twelvei=cmmi12
\font\twelvesy=cmsy10 scaled\magstep1
\font\teni=cmmi10
\font\tensy=cmsy10
\font\seveni=cmmi7
\font\sevensy=cmsy7
\font\it=cmti10 scaled \magstep3
\font\bf=cmb10 scaled \magstep3
\font\sl=cmsl10 scaled \magstep3
\textfont0= \eighteenrm \scriptfont0=\twelverm
\scriptscriptfont0=\tenrm
\def\rm{\fam0 \eighteenrm}
\textfont1=\eighteeni  \scriptfont1=\twelvei
\scriptscriptfont1=\teni
\def\mit{\fam1 } \def\oldstyle{\fam1 \eighteeni}
\textfont2=\eighteensy \scriptfont2=\tensy
\scriptscriptfont2=\sevensy
\def\doublespace{\baselineskip=30pt\lineskip=0pt
\lineskiplimit=-5pt}
\def\singlespace{\baselineskip=20pt\lineskip=0pt
\lineskiplimit=-5pt}
\def\oneandahalf{\baselineskip=25pt\lineskip=0pt
\lineskiplimit=-5pt}
\def\deg{^{\raise2pt\hbox{$\circ$}}}}


\def\reidel{\vsize=21.6cm\hsize15.2cm\hoffset.7in\voffset.65in
\twelvepoint\rm
\baselineskip=13pt\lineskip=0pt\lineskiplimit=-5pt
\parindent=12pt
\nopagenumbers
\null\vskip.5in
\def\title##1\par{{\leftline{\bf##1}}}
\def\author##1\par##2\par##3\par##4\par{{\moveright.85in
\hbox{\vbox{\parindent=0pt\parskip=0pt\vskip.6in
\noindent##1\hfill\break
{\it##2}\hfill\break
{\it##3}\hfill\break
{\it##4}\hfill\break}}\leftskip=0pt}}
\def\abstract##1\par{\vskip-.1in{\midinsert
\tenpoint\rm
\baselineskip=11pt\lineskiplimit=-5pt
\lineskip=0pt
\noindent ABSTRACT. ##1\endinsert}\vskip-.18in}
\def\section##1\par{\vskip26pt\noindent{\bf##1}\vskip13pt\noindent}
\def\subsection##1\par{\vskip13pt\noindent{\tenrm##1}\vskip13pt\noindent}
\def\subsec##1\par{\noindent{\tenrm##1}\vskip13pt\noindent}
\def\subsubsection##1..##2.{\vskip13pt\noindent{\rm##1}.\ {\it##2}. }
\def\subsubsec##1..##2.{\noindent{\rm##1}.\ {\it##2}. }
}

\def\cl{\centerline}

\def\clap#1#2{{\setbox0=\hbox{#1}\copy0\kern-0.5\wd0\setbox0=\hbox{#2}%
\kern-0.5\wd0\copy0}}

\def\eno{\the\eqano\global\advance\eqano by 1}

\def\gap{\mathrel{\hbox{\raise0.3ex\hbox{$>$}\kern-0.8em\lower0.8ex\hbox{
$\sim$}}}}

\def\iino{\the\iitemno\global\advance\iitemno by 1 }
\def\ino{\the\itemno\global\advance\itemno by 1 }

\def\lap{\mathrel{\hbox{\raise0.3ex\hbox{$<$}\kern-0.75em\lower0.8ex\hbox{
$\sim$}}}}

\def\microns{ifmode \mu m \else $\mu$m\if}

\def\pno{\the\pageno}

\def\ref{\noindent\hangindent=1.0in\hangafter=1}

\def\sno{\the\secno}
\def\ssno{\the\subsecno}

\def\subsection #1{\subsecno=#1}

\def\eg{{\it e.g.},}

\beginpreprint

\singlespace

\cl{\bf MULTIWAVELENGTH MONITORING OF THE BL LAC OBJECT
PKS~2155$-$304}

\vfill

\cl{\bf I.\ THE IUE CAMPAIGN}

\vfill

\cl{\it C.\ M.\ Urry,$^{1,2}$ L.\ Maraschi,$^{2,3}$ R.\ Edelson,$^4$
A.\ Koratkar,$^1$ J.\ Krolik,$^5$}

\cl{\it G.\ Madejski,$^{4,6}$ E.\ Pian,$^7$ G.\ Pike,$^4$ G.\
Reichert,$^{4,6}$ A.\ Treves,$^7$ W.\ Wamsteker,$^8$ R.\ Bohlin,$^1$}

\cl{\it J.\ Bregman,$^9$ W.\ Brinkmann,$^{10}$ L.\ Chiappetti,$^{11}$ T.\
Courvoisier,$^{12}$ A.\ V.\ Filippenko,$^{13}$}

\cl{\it H.\ Fink,$^{10}$ I.\ M.\ George,$^{4}$ Y.\ Kondo,$^{14}$ P.\ G.\
Martin,$^{15,16}$ H.\ R.\ Miller,$^{17}$ P.\ O'Brien,$^{18}$}

\cl{\it J.\ M.\ Shull,$^{19}$ M.\ Sitko,$^{20}$ A.\ E.\ Szymkowiak,$^4$ G.\
Tagliaferri,$^{21}$ S.\ Wagner,$^{22}$ R.\ Warwick,$^{23}$}

\vfill

\tobe{10 July 1993}{Astrophysical Journal}

\vfill

\recacc 1 December 1992; 7 January 1993

\vfill

\item{$^1$}Space Telescope Science Institute, 3700 San Martin Drive,
Baltimore, MD 21218

\item{$^2$}Guest Observer with the International Ultraviolet Explorer

\item{$^3$}Department of Physics, University of Milan, via Celoria 16,
I-20133 Milan, Italy

\item{$^4$}Laboratory for High Energy Astrophysics, Code 660, NASA/GSFC,
Greenbelt, MD 20771

\item{$^5$}The Johns Hopkins University, Department of Physics and
Astronomy, Baltimore, MD 21218

\item{$^6$}Universities Space Research Association, Code 610.3, NASA/GSFC,
Greenbelt, MD 20771

\item{$^7$}SISSA/ISAS International School for Advanced Studies, Trieste, Italy

\item{$^8$}ESA {\it IUE} Observatory, P.O. Box 50727, 28080 Madrid, Spain

\item{$^9$}Department of Astronomy, Dennison Bldg., University of Michigan,
Ann Arbor, MI 48109

\item{$^{10}$}MPE, Giessenbachstrasse, D-8046 Garching bei M\"unchen, Germany

\item{$^{11}$}Istituto di Fisica Cosmica CNR, via Bassini 15,
I-20133 Milan, Italy

\item{$^{12}$}Observatory of Geneva, Ch-1290 Sauverny, Switzerland

\item{$^{13}$}Department of Astronomy, University of California, Berkeley,
CA 94720

\item{$^{14}$}Laboratory for Astronomy and Solar Physics, NASA/GSFC,
Greenbelt, MD 20771

\item{$^{15}$}Canadian Institute for Theoretical Astrophysics, University
of Toronto, Ontario M5S 1A7

\item{$^{16}$}Also, Theoretical Astrophysics, 130-33, Caltech, Pasadena,
CA 91125

\item{$^{17}$}Department of Physics, Georgia State University,
Atlanta, GA 30303

\item{$^{18}$}Department of Physics and Astronomy, University College London,
Gower Street, London WC1E 6BT, England

\item{$^{19}$}JILA, University of Colorado, and National Institute of
Standards and Technology, Campus Box 440, Boulder, CO 80309

\item{$^{20}$}Department of Physics, University of Cincinnati, 210
Braunstein M1 11, Cincinnati, OH 45221

\item{$^{21}$}ESTEC, Space Science Department, Astrophysics Division,
Postbus 299, NL-2200 AG Noordwijk, Netherlands

\item{$^{22}$}Landessternwarte Heidelberg-K\"onigstuhl, K\"onigstuhl,
D-6900 Heidelberg 1, Germany

\item{$^{23}$}Department of Physics, University of Leicester,
University Road, Leicester LE1 7RH, England

\endtitlepage

\abstract{Daily monitoring of PKS~2155$-$304 with the IUE satellite
throughout November~1991 has revealed dramatic, large-amplitude, rapid
variations in the ultraviolet flux of this BL~Lac object. Many smaller, rapid
flares are superimposed on a general doubling of the intensity. During the
five-day period when sampling was roughly continuous, the rapid flaring had an
apparent quasi-periodic nature, with peaks repeating every
$\sim0.7$~days. The short- and long-wavelength ultraviolet light curves are
well correlated with each other, and with the optical light curve deduced from
the Fine Error Sensor (FES) on IUE. The formal lag is zero but the
cross-correlation is asymmetric in the sense that the shorter wavelength
emission leads the longer. The ultraviolet spectral shape varies a small but
significant amount. The correlation between spectral shape and intensity is
complicated; an increase in intensity is associated with spectral hardening,
but
lags behind the spectral change by $\sim1$~day. The sign of the correlation is
consistent with the nonthermal acceleration processes expected in relativistic
plasmas, so that the present results are consistent with relativistic jet
models, which can also account for quasi-periodic flaring. In contrast,
currently proposed accretion disk models are strongly ruled out by the
simultaneous optical and ultraviolet variability.}

\section 1.\ INTRODUCTION

The most puzzling aspect of Active Galactic Nuclei (AGN)
has always been their high power output coupled with the
small emission region inferred from rapid variability.
The characteristics shared by the
most rapidly variable objects, BL~Lac objects and Optically
Violently Variable (OVV) quasars,
collectively called ``blazars,''
such as high (and variable) polarization, compact radio structure,
a smooth continuum spectrum from radio through soft X-ray wavelengths,
and superluminal motion,
may owe their origin to a relativistic jet (Blandford~\&
Rees~1978).
The unreasonably high inferred radio brightness
temperatures ($T \gg 10^{12}$ K; Quirrenbach~\etal1989) and flare quotients
in excess of the Eddington-limited value assuming accretion efficiency
$\eta$ (Fabian~1979),
$\Delta L / \Delta t > 2 \times 10^{42} ~ \eta$~ergs~s$^{-2}$
(\eg~Feigelson~\etal1986; Morini~\etal1986),
often exhibited by blazars,
can most easily be explained by relativistic effects.
The quantity $\Delta L/\Delta t$ is proportional to $\delta^5$, where
$\delta = (\gamma [1-\beta \cos \theta])^{-1}$
is the kinematic Doppler factor describing relativistic motion
with Lorentz factor $\gamma$ (velocity $\beta$) at an angle $\theta$
to the line of sight.

In the last decade, considerable progress has been made interpreting the
broad-band spectra of blazars in terms of models of inhomogeneous
relativistic jets (Marscher~1980;
K\"onigl~1981; Ghisellini,
Maraschi~\& Treves~1985; Worrall~\etal1986; Hutter~\& Mufson~1986;
George, Warwick~\& Bromage~1988).
These models have been very successful, in the sense that with a minimal
number of parameters they usually fit the continuum spectrum over nearly
10 decades in wavelength. Unfortunately, the parameters of the model are
rarely well-determined because a variety of assumptions can produce acceptable
fits for a large volume of parameter space.
The degeneracy of
multiple model solutions vanishes or is greatly reduced when variability
information is added because the change in spectrum with intensity
is a strong diagnostic of the emission process (\eg~George~\etal1988;
Mufson~\etal1990). However,
the sampling available to date---only a few spectra, spaced far apart in
time---has been sparse, uneven, and inadequate.

Inhomogeneous jet models are strongly supported by radio and optical
observations, but the situation at higher frequencies is less clear.
Alter-\break natives include a two-temperature accretion
disk-\break model
(Wandel~\& Urry~1991) and
gravitational\break micro-lensing of background quasars
(Ostriker~\& Vietri~1985, 1990;
Stickel, Fried~\& K\"uhr~1988; Schneider~\& Weiss~1987).
The disk and jet models differ most at ultraviolet through X-ray
wavelengths, where the spectral curvature is greatest;
among jet models there are differences in how the ultraviolet and
X-ray emission are related (\eg~Ghisellini, Maraschi~\& Treves~1985).
Simultaneous, multiwavelength, well-sampled light curves are ideal for testing
the models. The amplitude and rapidity of variability in
most blazars increases with
decreasing wavelength, so the chances of successfully observing significant
variations at ultraviolet and soft X-ray wavelengths are high.
As other monitoring programs (with the different goal of mapping the
broad-emission-line regions in non-blazar AGN)
have clearly demonstrated (\eg~Clavel~\etal1991),
the IUE observatory is very well suited to regular monitoring of AGN
because of its ease of scheduling, efficient geosynchronous orbit,
precise photometric calibration and stability,
and broad wavelength coverage.

\eject

We designed a monitoring program
that would produce high-quality light curves in several bands, including
unprecedented spectral coverage
in the ultraviolet, extreme ultraviolet,
and soft-X-ray from the combination of IUE, the Rosat
Wide Field Camera (WFC), and the Rosat Position Sensitive Proportional Counter.
The object selected for our study, PKS~2155$-$304, is one of the brightest
extragalactic objects in the ultraviolet and X-ray sky.
Like most BL~Lac objects, PKS~2155$-$304 has no strong emission features;
the reported redshift of $z=0.117$
(Bowyer~\etal1984) is probably due to a galaxy displaced $\sim4$~arcsec
from the BL~Lac, but
a redshift of $\sim0.1$ can be inferred from imaging of the
host galaxy (Falomo~\etal1991).
PKS~2155$-$304 has previously
been observed to be highly variable at both ultraviolet
(Maraschi~\etal1986; Urry~\etal1988; Edelson~\etal1991) and X-ray
(Snyder~\etal1980; Sembay~\etal1992) wavelengths,
with some evidence of quasi-simultaneous variability in those bands,
albeit with smaller amplitudes and longer time scales at the longer wavelengths
(Treves~\etal1989).
The soft X-ray spectrum is steep, and can be connected smoothly to the
ultraviolet spectrum,
implying that the ultraviolet
and X-ray emission mechanisms may be related. This motivated the first
application of relativistic jet
(Urry~\& Mushotzky~1982) and accretion disk (Wandel~\& Urry~1991)
models to the continuum emission from PKS~2155$-$304.

In 1990, proposals to observe PKS~2155$-$304
were submitted to IUE (one to NASA, one to ESA)
and to Rosat (a U.S.\ proposal for month-long daily monitoring which was
unsuccessful and a German
proposal for intensive monitoring over a few days).
PKS~2155$-$304 is bright enough to be observed easily in a half IUE shift
(four~hours) or once per Rosat orbit (roughly 2000~seconds).
This paper describes the IUE data set, which is of unprecedented quality.
Associated observations at other wavelengths are being reported separately,
including month-long ground-based optical, infrared, and radio monitoring
(Smith~\etal1992; Courvoisier~\etal1992), quasi-continuous four-day Rosat
observations (Brinkmann~\etal1992), and multiwavelength cross-correlations
(Edelson~\etal1993).
The IUE observations and data analysis are described in
\S~2 of this paper, followed by the results in \S~3.
The implications of the ultraviolet variability for models of blazars
are discussed in \S~4 and the conclusions are summarized in \S~5.

\vskip1in

\section 2.\ OBSERVATIONS AND DATA ANALYSIS

\subsection 2.1 Observing Strategy

The Rosat spacecraft constraints restricted observing to a 32-day period
from 27~October~1991 to 28~November~1991, so the monitoring campaign
was planned for that time. (In the end, Rosat monitored the source only for
a few days in the middle of the month; Brinkmann~\etal1992.)
The variability time scales of BL~Lac objects in general, and even
this best-studied object PKS~2155$-$304 in particular, were not well
measured previously, so the observing plan bracketed a range of time scales.
In order to measure moderate time scale variations (days to a week) we
scheduled at least one half IUE shift daily from
1~November to 29~November (except on November~8, due to a scheduling conflict).
In order to study short-term variability,
4.6~days in the middle of the campaign (10.7--15.3 November)
were devoted to nearly continuous coverage using $\sim3$ shifts per day.
A log of the 201 IUE observations is given in Table~1.

The short-wavelength (SWP) and long-wave-\break length (LWP) IUE cameras were
exposed alternately, with nominal integration
times of 55 and 25~minutes, respectively. This allowed us to get two pairs of
spectra during each half IUE shift, in the absence of any operational problems.
Just before each SWP or LWP exposure, counts from the Fine Error Sensor
(FES), the optical monitor on IUE, were measured on target and on background.
During the continuous observing period, the SWP/LWP/FES observing cycle was
slaved to the 95.8-minute Rosat orbital period so that the
IUE and Rosat observations would be locked in phase, thus greatly simplifying
the cross-correlation between ultraviolet and X-ray light curves.
As a result of this rigid schedule, some IUE exposures had to be longer
or shorter than the nominal exposure times,
typically by a few minutes but occasionally by much more.
Exposure times for each IUE image are listed in Table~1.

\subsection 2.2 Optical Calibration

The FES counts were converted to optical magnitudes using the recently
developed algorithm of Perez~\& Loomis~(1991), which takes into account the
background due to scattered light. PKS~2155$-$304 was assumed to have color
$B-V=0.26$~mag throughout the month, which is the mean value measured
contemporaneously with ground-based optical telescopes (it did not change
much during the monitoring period; Smith~\etal1992).
In any case, the conversion from FES counts to V magnitude is not terribly
sensitive to the color; for example, using $B-V=0.5$~mag
would increase V by about 0.06~mag.
Reddening corrections are minimal at these wavelengths (well within the
FES accuracy), and so were ignored.
The FES-derived V~magnitudes are given in Table~1
and the optical light curve is discussed in \S~3.1.

\subsection 2.3 Ultraviolet Spectral Extraction

Spectra were extracted from each of the 201 IUE images using the
Slit-Weighted Extraction Technique (SWET) of Kinney, Bohlin~\& Neill~(1991).
Note that SWET-extracted spectra are the standard output for low-dispersion
images in the Final IUE Archive that is now being created
jointly by NASA, ESA, and SERC (the British Science and Engineering Research
Council).
The current IUESIPS software still uses a boxcar extraction, but the
SWET software is publicly available through the IUE Regional Data Analysis
Facility and has been incorporated into a number of processing pipelines,
including our own.

Details of the SWET procedure can be found in Kinney~\etal(1991), and we
merely summarize the major points here.
Empirical spline fits are made to the
cross-dispersion point spread function (PSF) along the
spectrum, which constrains the PSF to vary smoothly in the dispersion
direction.  This imposition of information contributes to an improvement
in the signal-to-noise ratio of the extracted spectrum relative to IUESIPS,
while flux is still conserved.
The flux at each wavelength is then determined by fitting the empirically
determined profile, sample by sample, weighted according to
a noise model determined from studies of hundreds of unrelated IUE images.
Discrepant points are eliminated, which automatically removes most of the
cosmic rays, with the exception of those that fall exactly at
the center of the cross-dispersion profile.
An uncertainty is associated with each flux value based on the noise model.
Extensive comparisons of SWET-extracted
and boxcar-extracted spectra show no systematic disagreement between the two
(Kinney~\etal1991).

There is another well-known slit-weighted technique, the Gaussian extraction
or GEX method\break (Urry~\& Reichert~1988), which assumes the cross-dispersion
profile is a Gaussian. Because of this additional constraint,
GEX gives a somewhat better signal-to-noise ratio than SWET for very low
signal-to-noise data, but for well-exposed spectra, systematic problems at
the level of a few percent can occur
because the true PSF is not precisely Gaussian.
In the present case, the spectra are generally well-exposed, so that
the SWET method is slightly preferred to the automatic
GEX algorithms currently
available, which in any case do not produce an error vector.
Nonetheless, we have also extracted
all the spectra using GEX, and compared this to the SWET results.
Although there are small systematic differences between spectra extracted
with the two extraction methods, the results (which depend on fitted
fluxes; see \S~2.5) are not affected significantly (see \S~3.4).

\subsection 2.4 Ultraviolet Spectral Corrections

The extracted net fluxes were converted to absolute flux using the
IUE calibration of Bohlin~\etal(1990), which produces absolute
fluxes that are up to 10\%
smaller than the white dwarf calibration (Finley~\etal1993)
being used for the IUE Final Archive.
A correction was made for degradation in the
SWP sensitivity (Bohlin~\& Grillmair~1988, as updated through 1989.36
by Bohlin, private communication).
The extrapolation to 1992 introduces some noise on the 5~\AA\ scale
but should not cause absolute flux errors of more than 1\% in broad bands.
No sensitivity correction was made to the LWP net flux.

We then considered a reddening correction.
The column density of hydrogen through
the interstellar medium of our Galaxy toward PKS~2155$-$304,
as measured with the $2^\circ \times 3^\circ$-beam
21-cm survey of Stark~\etal(1992),
is $N_{HI} = 1.78 \times 10^{20}$ atoms~cm$^{-2}$.
Using an IUE-based average conversion of
$\log N_{HI} /$\break $E(B-V) = 21.72 \, \pm \, 0.26$~cm$^{-2}$~mag$^{-1}$
(Shull~\& Van Steenberg~1985),
this corresponds to $E(B-V)\sim0.034$~mag, with large uncertainties
(discussed in detail in \S~3.4).
The actual line-of-sight column density could be different due to
small-scale, high-latitude fluctuations (\eg~Elvis, Lockman~\&
Wilkes~1989), but the soft X-ray spectrum of PKS~2155$-$304
is consistent with this column density of absorbing cool gas
(\eg~Canizares~\& Kruper~1984; Madejski~1985),
as are results from the Hopkins Ultraviolet Telescope,
which is sensitive down to 912~\AA\ (J.~Kruk, private communication).
These results all indicate there is
little or no internal reddening in PKS~2155$-$304.

The IUE spectra were dereddened using $E(B-V)=0.034$~mag and
the average Galactic curve from Seaton~(1979). The dereddening correction
has often been ignored by previous IUE observers of PKS~2155$-$304,
ourselves included, but it makes a significant difference.
The dereddened flux is 31\% greater at 1400~\AA\ and 21\%
greater at 2800~\AA\ than the observed flux, and the fitted energy spectral
index is typically 0.07 flatter in the SWP and 0.32 flatter in the
LWP. Because the amount of dereddening\dag\break
is uncertain, the
absolute ultraviolet luminosity is uncertain, but this remains true whether
or not the dereddening correction is applied.

Since we are interested in the source properties
unmodified by the accident of transmission through our Galaxy,
further discussion focuses on the dereddened spectra,
keeping in mind that the amount of reddening is uncertain.
The accuracy of the assumed Galactic reddening is
considered further in \S~3.4.

\subsection 2.5 Ultraviolet Spectral Fitting

Using an iterative, chi-squared minimization fitting routine,
the dereddened IUE spectra were fitted to a simple power-law model
of the form

$$ {F_\lambda}  = { b_1 {\left ( {{\lambda}\over{\lambda_0}} \right )
}^{b_2} } ~~.\eqno(1)$$

\noindent
The fit parameters are the normalization, $b_1$, at fiducial wavelength
$\lambda_0$, and the slope, $b_2$.
The definition of power-law slope
seen most often in the literature is the energy index, $\alpha$, where
$F_\nu \propto \nu^{-\alpha}$, which is related to $b_2$ via
$\alpha = 2 + b_2$. The results are given here in terms of the energy index,
$\alpha$, but the fitting was done in wavelength space, with no resampling
to frequency space.

Two sample fits to the data,
chosen to be representative of the median intensities and the mean
$\chi_\nu^2$-values for the SWP and LWP samples as a whole,
are shown in Figure~1; the SWET error vectors are
plotted below each spectrum. (In order to show the raw extracted spectra,
the data and fits in Fig.~1 are not dereddened.)
The wavelength ranges over which the data were fitted were
1230--1950~\AA\ for the SWP camera (which excludes
the geo-coronal Lyman-$\alpha$ region) and 2100--3100 for the LWP camera.
Wavelength regions affected by SWP camera artifacts (Crenshaw,
Bruegman~\& Norman~1990), at 1277--1281~\AA,
1286--1290~\AA, and 1660--1666~\AA, were excluded, as was
the region 1470--1540~\AA, in which unusual features
were apparent in many of the spectra.
These excluded regions are shown as light dotted lines in Fig.~1a.
Since\break
\smallskip
\hrule width 1.75in
\smallskip
\noindent \dag The effect of reddening is greatest at the shortest wavelengths,
but the change in spectral index is actually greater in the
LWP band than in the SWP band because the slope of the reddening
curve over the heavily weighted part of the LWP band is larger than
the slope in the SWP band.

\vskip1in

\noindent the artifacts in the LWP camera are of low contrast,
no spectral regions were excluded in those fits.
The power-law fit is clearly good, as is generally true
for all the spectra.

Results of the fits to dereddened spectra are given in Table~2.
The normalization
is at 1400~\AA\ for SWP spectra and 2800~\AA\ for LWP spectra
(as in Edelson~\etal1992).
These wavelengths are close to the flux-weighted means of each
band ($\sim1560$~\AA\ and $\sim2568$~\AA, respectively, for $\alpha =1$),
so that the uncertainty in the derived flux is small;
they are in regions of the cameras where the signal-to-noise ratio is good;
and they were chosen to be relatively far away from one another, increasing
the independence of the SWP and LWP flux measurements.
The reduced chi-squared values
in Table~2 were calculated using the raw SWET errors,
prior to the correction procedure described in the next section.

We also fitted combined spectra, which is to say pairs of
SWP and LWP spectra taken close together in time, spliced
together at 1978~\AA.
Power laws were fitted to the wavelength range 1230--3100~\AA,
excluding the same regions as in the SWP analysis above,
as well as 1900--2150~\AA\ (which has relatively large errors anyway).
The results are given in Table~3.
The advantage of fitting combined SWP-LWP spectra is that the longer baseline
in wavelength gives smaller uncertainties on the fitted flux and spectral
index; however, the mismatch between SWP and LWP spectra seen in
previous work (\eg~Urry~\etal1988; George~\etal1988)
and again here (see \S~3.3)
illustrates how the absolute values of fitted flux and spectral index
depend on the uncertain intercalibration of the two cameras.
Therefore, for absolute values of the flux and spectral index,
the estimates from fits to the individual spectra
($F_{1400}$, $F_{2800}$, $\alpha_{SWP}$, and $\alpha_{LWP}$, in
Table~2) are better than those from the combined fits
($\alpha_C$ and $F_C$, in Table~3).
We consider the latter quantities when evaluating
{\it changes} in spectral shape and intensity
because of the larger bandwidth for the power-law fits
(the combined spectral index is an effective hardness ratio
between the SWP and LWP cameras).

\subsection 2.6 Error Analysis

Estimating the error bars reliably is the key to the detection and evaluation
of variability. IUE spectra are dominated by systematic noise, including
well-known features at the wavelengths mentioned above.
A continuum estimate from direct measurement
would have a relatively large error bar (the variance in some interval
around that wavelength), including both local statistical noise and
fixed-pattern noise.
Like most BL~Lac objects, however, PKS~2155$-$304 has a smooth and
approximately featureless
spectrum which is well fitted by a simple power-law model.
The flux calculated from a fitted power law
gives a much smaller error bar than direct measurement
because information from the full band is used.

The initial estimate of the uncertainty in the flux measurement comes
directly from propagated uncertainties on fit parameters. The latter
are derived from the error matrix calculated in the least-squares
fitting procedure (\eg~Bevington~1969),
which in turn depends linearly on the SWET errors used to weight the
input data points.
Thus, the estimated uncertainty in the flux measurement
scales linearly with any modification to the SWET error.

While the relative sizes of the SWET errors make sense,
the chi-squared distribution for
the 201 spectral fits is far from acceptable. The mean reduced chi-squared
values for the SWP and LWP fits are $\langle \chi_\nu^2 \rangle = 3.76$
and $\langle \chi_\nu^2 \rangle = 2.42$, respectively,
where the number of degrees of freedom is taken as the number of points
fitted in the spectrum (539 and 535, respectively)
minus the number of fit parameters (2).
The probability of exceeding either of these values given the number
of degrees of freedom is vanishingly small.
The power-law model appears to be a good fit to the data, and there
are no systematic trends in the residuals that would indicate
a different model might be preferred, so
either the SWET errors are underestimated or
the number of degrees of freedom (the number of independent points in each
spectrum) is overestimated.

Previous authors (\eg~Clavel~\etal1991) have suggested normalizing the
error vector for a given spectrum by the square-root of the
reduced chi-squared
value for that fit. In our view this is not the best approach, as it
makes the $\chi_\nu^2$ distribution for all the fits look like a delta-function
when in fact the $\chi_\nu^2$ values should be distributed as chi-squared.
(It might be the best approach in the case of a single measurement.)
Here we have a large number of measurements (201),
so we imposed the condition that the mean of the reduced-chi-squared
distribution be 1, as is expected for this many
degrees of freedom. (The 102 SWP spectra and 99 LWP spectra were handled
separately; characteristics of the cameras such as graininess and
fixed-pattern noise are quite different, so there is no reason to expect
the corrections to the error vectors to be the same in the two cases.)

\vskip1in

Scaling the SWET errors by 1.94 for the SWP spectra and 1.56 for the LWP
spectra gave normal chi-squared distributions.  Not only were the mean values
1, by definition, but a Kolmogorov-Smirnov (K-S)
test showed no difference between the observed
distributions and the expected $P(\chi_\nu^2)$.
(The cumulative $\chi_\nu^2$-distributions are shown in Fig.~2.)
The uncertainties in the fitted fluxes were thus increased by those factors.
A similar procedure was followed for the combined fits,
where the mean reduced chi-squared value using the uncorrected sigmas
was $\langle \chi_\nu^2 \rangle = 3.42$. In this case, however, although
the mean of the corrected distribution was adjusted to be 1, a K-S test
gave a low probability ($4.7 \times 10^{-4}$) that the observed distribution
was drawn from a normal chi-squared distribution. Thinking this might be
related to the different camera characteristics, we re-fit the combined
spectra normalizing SWP errors by 1.94 and LWP errors by 1.56. The resulting
reduced-chi-squared distribution had a mean of 1.11,
so we renormalized the errors
by $1.05=\sqrt{1.11}$. The final $\chi_\nu^2$-distribution, with mean equal
to 1, was still incompatible with the expected distribution
($P<3.9\times10^{-5}$ according to a K-S test),
probably because of the mismatch between SWP and LWP spectra.
This reinforces our belief that the individual fits are better for measuring
fluxes or spectral indices,
while the combined fits may be more sensitive
(due to the broader bandwidth) to trends in either flux or spectral index.

\def\ie{{\it i.e.},}

There can still be residual errors in measured quantities, such as
spectral index or fiducial fluxes,
associated with the reproducibility
from one spectrum to the next. These must be included when
evaluating light curves. IUE fluxes of standard stars are
reproducible at the $\sim1.25$\% level (Bohlin~1988), so errors of
this magnitude were added in quadrature to the internal flux errors
estimated above (as in Edelson~\etal1991).
In general, the internal photometric error is considerably smaller
than 1.25\%, unless the exposure time is unusually short.
Estimating the repeatability of the spectral index measurements was more
complicated.  In their study,
Clavel~\etal(1991) estimated residual errors in measured quantities by
comparing close pairs of spectra and assuming
no variability on short time scales.
If the estimated errors were really 1~$\sigma$ errors,
the difference in measured values divided by the estimated error (the
quadrature sum of the two individual error estimates) should be
normally distributed, \ie~a Gaussian with zero mean and unit dispersion.
A variance larger than one, Clavel~\etal argued, meant the uncertainties
on that measured quantity should be increased by just that factor.
Since PKS~2155$-$304
is much more rapidly variable than their target, the Seyfert galaxy NGC~5548,
this procedure was not appropriate for our flux measurement, hence the 1.25\%
photometric correction adopted as described above.
However, it was the best alternative for
evaluating the uncertainty on spectral index, particularly as the spectral
index is certainly less variable than the flux (see \S 3.2).
Adjacent spectra taken closer in time than 5~hours were compared;
this included 75 pairs of SWP and 69 pairs of LWP spectra.
The variances in the distributions of normalized errors for short- and
long-wavelength spectral indices were 1.94 and 2.41, respectively,
so the error estimates in $\alpha_{SWP}$ and $\alpha_{LWP}$ were increased
by those factors. The distributions of normalized, scaled errors were
then consistent ($\apgt$50\%
probability according to a K-S test)
with the normal Gaussian distribution expected.
This procedure was repeated for combined spectra, comparing 78
pairs of adjacent, combined SWP-LWP spectra,
taken within 5~hours of one another.
Here, the scale factor for the combined
spectral index was 2.18, and the final distribution was consistent with
a Gaussian ($\sim59$\% probability, according to a K-S test).

The estimation of errors in the flux and spectral index, in summary,
involves several steps. First, the SWET error vector is propagated through
the fitting procedure to get initial uncertainties for the parameters
of the power-law fit, $b_1$ and $b_2$.
Next, uncertainties in both $b_1$ and $b_2$ are increased by a factor
equal to the square root of the mean of
the reduced chi-squared distribution, $\sqrt{\langle \chi_\nu^2 \rangle}$,
where $\langle \chi_\nu^2 \rangle = 3.76$ for the SWP and
$\langle \chi_\nu^2 \rangle = 2.42$ for the LWP.
This is effectively scaling up the SWET error vector so that
$\langle \chi_\nu^2 \rangle = 1$ in both cases.
The final error estimate on the flux at $\lambda_0$ is equal to the
quadrature sum of the scaled error on $b_1$ and 1.25\% of $b_1$.
The final error estimate on the spectral index was derived by
increasing the scaled error on $b_2$ by
1.94 for the SWP, 2.41 for the LWP, and 2.18 for the combined SWP-LWP fits,
so that it represents the 1~$\sigma$ error for the observed differences
between adjacent measurements of $\alpha$. Since this assumes no
intrinsic variation between adjacent measurements, it is if anything an
overestimate of the error on the spectral index.

As a check, we did look at the distribution of normalized errors in the
fluxes, using the uncertainties calculated as described above.
In all three cases (SWP, LWP, and combined SWP-LWP values),
the distributions were consistent with the expected Gaussian
distributions (62\%, 13\%, and 38\% probability, respectively,
according to a K-S test), indicating that our error estimate
for the flux is at worst an overestimate.

A procedure similar to that used for determining the uncertainty in
spectral indices was
used to estimate the global mean uncertainty in the
FES-derived magnitude. Using a trial value for the uncertainty,
we generated the normalized error distribution for pairs of
adjacent measurements separated by less than 5~hours,
looking for a value that would give a variance equal to 1.
The normalized error distribution for $\Delta V = 0.08$~mag had
variance equal to one and
was consistent (97\% probability according to a K-S test)
with a Gaussian distribution centered on zero.
It is also commensurate with previous estimates
($\Delta V\sim0.07$~mag) of the accuracy of the FES
(Holm~\& Crabb~1979; Barylak, Wasatonic~\& Imhoff~1984).
This is therefore a good estimate of the mean
uncertainty in the FES magnitudes, but it does not account
for possible infrequent, anomalous FES measurements, for
systematic trends in FES errors with FES count rate, or for any
systematic offset from other optical measurements
(\eg~Smith~\etal1992; Courvoisier~\etal1992),
such as might be due to incorrect color corrections.

\section 3.\ RESULTS

\subsection 3.1 Ultraviolet and Optical Light Curves

The monitoring campaign was a great success. Our BL~Lac object
cooperated nicely, increasing in intensity by a factor of 2 over the month
to roughly its historical maximum brightness.
The light curves for the full month and for the central period
are shown in Figure~3. During the intensive monitoring,
the ultraviolet flux varied by $\sim30$\% in several distinct flares that
are well-sampled apart from a possible dip during
the 7-hour gap on 11~November. The width of these rapid flares is roughly
half a day; if we define an exponential variability time scale as
$t_{var}=(d\, \ln F/dt)^{-1}$,
then values for these flares are less than 2~days.
Such fast ultraviolet variations have been detected previously in only
two blazars, PKS~2155$-$304 and Mrk~421, while slightly longer and/or lower
amplitude flares have been seen in another three (3C~279, OI~158, and OD~26;
Edelson~1992).
The ultraviolet flares of PKS~2155$-$304 show no discernible asymmetry in time
according to a $dI/I$ test (Wagner~\& Witzel~1992).
The depth of the dip during 12~November is unclear. The low SWP point
comes from a very short exposure of only 13~minutes---most SWP exposures
lasted 55~minutes, and the next shortest exposure was 25~minutes---and
no corresponding dip is seen in the LWP light curve,
so the SWP point should be considered uncertain (see \S~3.4).

Based on the rapid flaring seen during the intensive monitoring,
it appears we probably failed to sample the fastest-time-scale flares
during the rest of the month, although the overall doubling of the
flux is well-sampled.
The fractional variability is comparable for the SWP and LWP bands: both light
curves show a doubling of flux, and in both bands the variance is about
15\% of the mean flux. This is in contrast to the historical trends in other
blazars, where long-term IUE monitoring indicates the SWP flux is more variable
than the LWP flux (Edelson~1992).

The FES light curve, shown in Figure~4, shows the same trends
as the ultraviolet light curves, on both long and short time scales,
although the larger error bars mean the variations are less well defined.
Although the light curve is given as V-magnitude (from the algorithm
used to convert FES counts to optical magnitudes; Perez~\&
Loomis~1991),
the FES sensitivity is slightly bluer than the V-band.

The SWP, LWP, and FES fluxes are all highly correlated, as shown
in the discrete cross-correlation functions (DCF; Edelson~\&
Krolik~1988, with additional minor modifications by Krolik, unpublished)
for SWP versus LWP and SWP versus FES (Figure~5). The peaks of both DCFs are
at zero lag,with an upper limit of $\aplt0.1$~days, but
both cross-correlation functions are asymmetric, in the sense of the
short-wavelength emission leading the longer wavelength emission.
These are not necessarily contradictory; see interpretation in \S~4.2.

The autocorrelation functions of the SWP, LWP, and FES light curves are shown
in Figure~6. The SWP and LWP give similar results, while the FES amplitude
is generally smaller because the relative errors are larger.
(For the SWP and LWP, the behavior of the autocorrelation functions suggests
the estimates of flux errors were about right.
Note that in the method of Edelson~\& Krolik~[1988], the
amplitude of the cross-correlation or autocorrelation function
depends on the binning, and hence the error bars, so that it can
be formally larger than unity.)
On long time scales, the autocorrelation functions suggest smooth ``red''
power spectra, with relatively more power on longer time scales.
However, on shorter time scales (using the data from
the well-sampled, 4.6-day intensive monitoring)
they are modulated with a period of $\sim0.7$~days, with the first
harmonic also seen at $\sim1.5$ days.
This quasi-periodic behavior can be seen going through
five cycles in the light curves (Fig.~3b), despite
the gap at 11.5~November.


The significance of this possible periodicity is unclear.
Were there only a few peaks ($\aplt3$),
it could almost certainly be dismissed as spurious (Press~1978).
Signals with red power spectra frequently show spurious
periodicities on time scales that are an appreciable fraction of
the length of the data stream
because there are too few large-amplitude components at low frequencies to
achieve the Gaussian behavior of the central limit theorem.
The fact that the apparent period is closer to $1/5$ of the intensively-sampled
stream, and that the light curve shows little power at other
frequencies, tempts us to believe that it may be real; but this would
only be a comfortable conclusion if there were 10 or 20 periods (so that
the central limit theorem would indeed apply).
The sampling in contemporaneous optical monitoring was not sufficient to
detect periods shorter than a few days (Smith~\etal1992).
This period is not present in extensive, well-sampled
optical data taken in 1988 (Carini~\& Miller~1992), which means,
since we now know the optical and ultraviolet are closely related,
that any periodicity, if real, is transitory.
Only a longer run of intensive monitoring can clearly confirm
or refute the stability of this periodicity.

\subsection 3.2 Ultraviolet Spectral Variability

A key point in the interpretation of the continuum emission from
PKS~2155$-$304
is the relationship between variability in intensity and spectral shape.
Previous monitoring has suggested that the spectral index varies little
during flares in intensity, but this observation was always restricted to
grossly undersampled light curves. The present campaign, with its
improved sampling, should be more sensitive to related
variations in flux and spectral index.
The variation of spectral index throughout the run is shown in
Figure~7.
According to the chi-squared statistic, the model of constant spectral
shape can be ruled out:
$\chi_\nu^2 = 4.68$ for the combined spectral index.
This is also seen in the SWP camera alone, where
$\chi_\nu^2 = 1.94$ ($P(\chi_\nu^2) = 5.9\times10^{-8}$) for
a constant fit to $\alpha_{SWP}$.
(The variation in the LWP spectral index is not significant,
$\chi_\nu^2 = 0.98$, because of the larger errors due to the low
signal-to-noise ratio below 2400~\AA.)

Plausible physical processes causing intensity variability, such as
acceleration of radiating particles near a shock, predict accompanying
changes in the spectrum of the emitted radiation.
For example, electron acceleration would cause
spectral hardening with increasing intensity.
Thus one might expect flux and spectral index to be inversely correlated.
Such correlations have been found previously for Mrk~421
(Ulrich~\etal1984) and OJ~287 (Maraschi~\etal1986), although these studies
referred to data taken over much longer time intervals.
In the present case, a non-parametric Spearman
Rank-Order Correlation test suggests an inverse correlation
between $F_{1400}$ and $\alpha_{SWP}$ ($P=9.2\times10^{-3}$ of
occurring by chance for the full data set;
$P=0.07$ for the intensive monitoring period only),
but not between
$F_{2800}$ and $\alpha_{LWP}$ ($P=0.87$ and $P=0.76$
for full and intensive periods, respectively) or
$F_{2000}$ and $\alpha_{C}$ ($P=0.41$ and $P=0.19$ for
full and intensive periods, respectively).
Figure~8a shows the scatter plot of the last of these,
with filled circles indicating data from the intensive monitoring period.

Previous authors found no correlation between flux and spectral index for
PKS~2155$-$304, but did find a correlation
between $\Delta F$ and $\Delta \alpha$, where the change is measured between
pairs of spectra taken close in time
(Maraschi~\etal1986; Urry~\etal1988; Edelson~1992).
In the present instance, in contrast,
change in spectral index and change in flux
are not strongly correlated;
Figure~8b shows the change in spectral index versus the change in flux
for the combined SWP-LWP spectral fits.
The probability of correlation between
$\Delta \alpha$ and $\Delta F$ for the intensive monitoring period
(measured from adjacent data points)
is significant only for the lower-quality LWP data ($P>99.9$\%).
For the full data stream, using the differences between daily averages,
$\Delta \alpha$ and $\Delta F$ are marginally anti-correlated for
the SWP data ($\sim96.7$\%).
In neither case are the combined SWP-LWP
data correlated or anti-correlated.
We conclude that there is no significant direct correlation between
instantaneous spectral shape and intensity,
or changes thereof, in the IUE data.

Given the importance of understanding spectral variability,
we explored a new tool for measuring the phenomenon.
We applied the Discrete Correlation Function (Edelson~\& Krolik~1988),
usually used to investigate flux-flux correlations,
to the cross-correlation between flux and spectral index
for the intensive monitoring period, where the light
curve is well-sampled. The results are shown in Figure~9
for the SWP and LWP fits (which track each other well).
There is clearly structure in this DCF. The negative lag
means that changes in the spectral index precede
changes in the intensity, and the negative correlation amplitude
means that spectral hardening (decreasing $\alpha$)
is associated with increasing intensity.
Thus, there appears to be some kind of
inverse correlation between spectral index
and intensity, but with a lag of one to two days.

\subsection 3.3 Ultraviolet Spectral Shape

The spectral shape in the ultraviolet band depends critically on the
dereddening correction. When no correction is applied, the LWP spectral
indices are systematically higher than the SWP spectral indices
(\eg~Edelson~1992).
This implies flattening of the spectrum toward shorter
wavelengths; that is, a ``concave-up'' shape in $\nu$-$F_\nu$ space,
in contrast to the overall ``concave-down'' shape of the radio through
soft-X-ray spectrum.
An excess at long wavelengths due to inclusion of
starlight is not expected, as the host galaxy for PKS~2155$-$304 detected
in deep optical imaging (Falomo~\etal1991) is too faint ($V\sim16.5$)
relative to the nuclear point source.
The spectral curvature goes away, however, when
a dereddening correction with $E(B-V)=0.034$~mag is applied.
(If we fit the entire ensemble of combined SWP-LWP spectra to a power-law model
leaving the reddening correction as a free parameter,
chi-squared is minimized for this value.)
Figure~10 shows the distributions of SWP and LWP spectral indices for three
assumed values of $E(B-V)$:~0.0~mag, 0.034~mag, and 0.05~mag.
With no correction, the
LWP spectra are steeper; with $E(B-V)=0.05$~mag, the SWP
spectra are steeper; and
$E(B-V)=0.034$~mag is roughly the value at which the SWP and LWP
spectral indices agree best.
Thus, while the amount of reddening remains uncertain (see \S~2.3),
the assumption that the ultraviolet spectrum is either a flat power law or
concave-down would indicate a minimum reddening of $E(B-V)\sim0.03$~mag.

Taking the reddening into account, then, there is no evidence for spectral
curvature in the ultraviolet band. The flatness of the UV spectral index
($\alpha < 1$)
means that peak of the luminosity emitted from this BL~Lac object is
in the far-ultraviolet, as noted by previous authors.

\subsection 3.4 Systematic Errors

The absolute calibrations and sensitivity corrections
are the largest sources of uncertainty in the absolute value of the flux.
The IUE calibrations
available---the one used in our pipeline (Bohlin~\etal1990),
the white dwarf calibration being used for the IUE Final Archive
(Finley~\etal1993),
and the standard IUESIPS calibration (Cassatella, Lloyd~\&
Riestra~1988)---result
in calibrated fluxes that differ by less than 10\%. The Bohlin
calibration tends to be lower than the IUESIPS calibration for the SWP
(the average ratio is 0.93).

The SWP sensitivity correction is roughly 1\% per year averaged over the
whole waveband. The extrapolation from 1988 is uncertain.
We have applied no sensitivity correction to the LWP, although one has been
published (Teays~\& Garhart~1990). Such a correction would
possibly alleviate the camera mismatch, as in our results
the SWP flux is too high relative to the LWP.

There are small systematic differences in the shapes of IUE spectra
extracted with GEX and with SWET. However, the ratio of the extracted spectra
averaged over the full waveband is close to 1.
The fitted spectral indices in the two wavebands differ systematically
(by $\Delta\alpha\sim0.1$ for the LWP and $\Delta\alpha\sim0.2$ for the SWP)
but the results of combined fits are essentially identical.
Therefore, the results are independent
of extraction method. (The advantage of the SWET technique, providing
independent error estimates for each data point, remains.)

As mentioned earlier, the amount of reddening along the line of sight to
PKS~2155$-$304 is uncertain.
With no dereddening, the absolute ultraviolet flux would be
$\sim20$--30\%
lower than the numbers quoted in Table~2 and the spectral shape would be
concave-up (\ie~there would be excess long-wavelength flux above the SWP
power law). There is no obvious explanation for an intrinsic spectrum of
this shape, whereas concave down spectra can result naturally from
synchrotron losses. The available evidence is generally consistent
with our assumed value of $E(B-V)=0.034$~mag, which would correspond to
an approximately flat (\ie~not concave-down or -up) spectrum,
but the uncertainties are large.
One might reduce the uncertainty in $N_{HI}$ due to small-scale fluctuations
in the interstellar medium by using PKS~2155$-$304 as a background radio source
at 21~cm, although that would introduce another uncertainty due to the
unknown spin temperature of the gas.
Another approach would be to estimate the extinction in the extreme
ultraviolet,
which is very sensitive to the H~I and He~I column densities
(particularly the latter), using the recently obtained EUVE spectrum
(Malina~\& Bowyer~1992). The total optical depth at 100 eV corresponding to
$N_{HI} = 1.78 \times 10^{20}$ atoms/cm$^2$, assuming He~I is 10\% by number,
is $\tau\sim11$, quite high considering that PKS~2155$-$304
is one of the two brightest extragalactic sources detected
with the Rosat WFC (Pounds~\etal1992), and was also detected
in the short-wavelength ($E\sim100$~eV) filter of EUVE
(Malina~\& Bowyer).

After our analysis was completed, we learned of an
unpublished 21-cm emission measurement by Lockman and Savage~(1993),
$N_{HI}=1.36 \times 10^{20}$ cm$^{-2}$,
made with a 21-arcminute beam and
careful attention to the antenna sensitivity pattern
(\eg~see description in Lockman, Jahoda~\& McCammon\break 1986).
This value for the Galactic hydrogen column density,
which is somewhat lower than the Stark~\etal(1992) value we used,
would corresponds to $E(B-V)=0.026$~mag and
would result in lower absolute fluxes than those given in Table~2,
by about 6\% for the SWP and 3\% for the LWP, or about
22\% and 18\% higher, respectively, than the observed fluxes.
The total optical depth at 100~eV for this column density would be $\tau\sim9$.

In any case, the uncertainty in the gas-to-dust ratio dominates the
uncertainties in the assumed $E(B-V)$. The scatter in the conversion
factor deduced by Shull~\& Van Steenberg~(1985), for example, gives a
1~$\sigma$ range for $E(B-V)$ of 0.014--0.047~mag.
In addition, we used the conversion
appropriate to the full sample of 205 stars; a conversion for
the 53 halo stars only,
$\log N_{HI} / E(B-V) = 21.83$~cm$^{-2}$~mag$^{-1}$,
gives $E(B-V)=0.026$~mag for the Stark~\etal(1992) value of
$N_{HI}$, or 0.020 mag for the Lockman~\& Savage value of $N_{HI}$,
with somewhat larger uncertainties.

Collectively, the uncertainties in absolute calibration,
sensitivity
corrections, and dereddening corrections
amount to 10--15\% uncertainties
in the measured absolute flux. However, since the same corrections are
applied to all the spectra, the relative fluxes are not affected by these
uncertainties.

Like many blazars, PKS~2155$-$304 can be highly polarized
(\eg~Smith~\etal1992; Allen~\etal1992).
Because the IUE spectrographs utilize gratings, spurious variability could
in principle be introduced by changes in polarization. During the monitoring
campaign, the U-band polarization of PKS~2155$-$304 was always $\aplt8$\%
(Smith~\etal1992), so that the effect would have been
no larger than the quoted uncertainties.

Finally, systematic errors arising in only a handful of spectra or FES
images are not taken into account by our error analysis procedure,
which assumes that the exposures have similar properties.
This is probably a reasonable assumption for most of the data,
which were obtained under similar conditions.
Exceptions include the four unusually short SWP exposures
(points circled in Figure~3), since short IUE exposures have been shown
to give systematically low fluxes (Walter~\& Courvoisier~1991).
The errors on quantities measured from these SWP exposures are almost
certainly underestimated.

\section 4.\ DISCUSSION

\subsection 4.1 Intensity Variability

Throughout our month-long observing campaign the ultraviolet and optical
flux of PKS~2155$-$304 increased by a factor of two.
The light curves are far from smooth, however,
with rapid lower-amplitude events superimposed.
During the intensive monitoring period
in the middle of the month,
there are a number of well-sampled, large-amplitude ($\sim30$\%) variations
which appear to have a quasi-periodic time scale of $\sim0.7$~days.
Whether this periodicity is real requires a longer data train with
equal or better sampling.
The exponential time scales associated with the most rapid events
are a few days or less;
the actual time in which flux was seen to double was approximately 10 days.
Similar fast, intraday variability, including the appearance of
quasi-periodicity, has been seen in optical observations
of BL~Lac objects (Quirrenbach~\etal1991).
Such short time scales are consistent with the predictions of the
standard relativistic jet and accretion disk models for BL~Lac objects
(\eg~K\"onigl~1981; Wandel~\& Urry~1991).

The possible periodicity at $\sim$15--20~hours is one of the most
intriguing yet ultimately frustrating results of the present campaign.
We have sufficient data that we cannot automatically dismiss
it as an artifact of the length of observation or of poor sampling;
on the other hand, we lack enough data to be confident of its reality.
During our observations, PKS~2155$-$3404 was in a particularly bright state,
as was the BL~Lac object OJ~287 when a possible periodicity was
detected in its optical light curve (Carrasco,
Dultzin-Hacyan~\& Cruz-Gonzalez~1985).
It is clearly important to obtain longer, high-quality light curves
of PKS~2155$-$304 in a bright state.

Meanwhile, keeping in mind the uncertainties,
we explored what such a period, if real, might mean.
One possible origin of a periodicity is orbital motion.
The nearest stable orbit around a black hole is
at a few ($n\apgt3$) Schwarzschild radii,
where $R_S = 2GM_{BH}/c^2$. Using Kepler's law and ignoring
relativistic effects, the central black hole mass is related to
is the orbital period, $T$,
by $M_{BH} = T c^3 / (4 \sqrt{2} \pi n^{3/2} G)$.
If the ultraviolet emission is not relativistically beamed,
then $M_{BH} < 1.3 \times 10^8~M_{\odot}$.
Relativistic beaming would shorten the observed time scale relative to
the true time scale, and so would allow higher masses
in this simple limit.

In the context of a relativistic jet picture,
the periodicity can be explained if the production of the
jet is closely coupled to the orbital dynamics.
For example, Camenzind~\& Krockenberger~(1992) considered jets that
are magnetically collimated winds emanating from the inner edge
of an accretion disk. In this model,
spiraling motion of plasma bubbles moving along magnetic field
lines in the jet causes quasi-periodic fluctuations in
observed intensity simply as a result of a smooth
variation in the Doppler factor.
The rotation period of the escaping plasma bubbles
is initially the Keplerian period of the inner disk,
increasing beyond the Alfven point, $R_A$, as $(R/R_A)^2$,
where $R$ is the perpendicular distance of the bubble from the jet axis.
The bubbles become collimated and are observed at $R>R_A$.
The observed period will be shorter by the factor
$(1-\beta \cos \theta$), where $\theta$ is the inclination of the jet,
due to a relativistic projection effect.

For PKS~2155$-$304 there are no firm estimates of either the Lorentz
factor or the inclination angle (cf.~Urry~\& Mushotzky~1982 and
Ghisellini~\etal1985),
but assuming $\gamma\sim5$, as appropriate
for one of the more active X-ray-selected BL~Lac objects
(Padovani~\& Urry~1990), and $\theta\aplt1/\gamma\sim10^\circ$,
the observed period of 0.7 days corresponds to
a true period of $1.7\times 10^6$~seconds. For
$R/R_A\sim10$ (Camenzind~\& Krockenberger~1992),
the Keplerian period at $5R_S$ is then $1.7\times10^4$~seconds,
and the black hole mass is $1.8\times10^7~M_\odot$.

The ultraviolet flux from PKS~2155$-$304 varies between
1.3 and $2.7\times 10^{-10}$~ergs~cm$^{-2}$ s$^{-1}$
(1200--3000~\AA).
For a redshift of $z\sim0.12$
(assuming $H_0=50$~km/s/Mpc, and isotropic emission),
this corresponds to an ultraviolet luminosity of
0.9 to 1.9$\times 10^{46}$~ergs~s$^{-1}$.
A 30\% change in intensity occurring over a day (for example,
around 13--14~November~1991) corresponds to
$\Delta L/\Delta t\sim5 \times 10^{40} $~ergs~s$^{-2}$,
which is only a factor of 4 below the fiducial limit for
Eddington-limited accretion with efficiency $\eta= 0.1$
(\S~1).
We estimate a bolometric correction of $\sim10$ for PKS~2155$-$304
using published multiwavelength data.
(Given the flat ultraviolet spectral index, $\alpha<1$,
the bolometric correction depends most
strongly on the detailed ultraviolet through soft-X-ray
spectrum, for which we use the data summarized by Wandel~\& Urry~1991.)
If the UV, extreme UV, and X-ray flux all vary comparably
(see Edelson~\etal1993),
the flaring monitored in November~1991 would in fact
have exceeded the limit, suggesting relativistic beaming is present.

\subsection 4.2 Spectral Variability

The spectral variability observed is intriguing, and not easily interpreted.
The fluxes within both IUE bands and the optical flux deduced from
the FES are all well correlated with no discernible lags on
time scales $\apgt3$~hours.
The cross-correlation functions are asymmetric, however, in the sense
that short-wavelength emission leads the long.
This may indicate that the emitting volumes are not significantly different
across the ultraviolet band, but that radiative losses are sufficiently
energy dependent that the short-wavelength flux decays faster than the
long-wavelength flux.

The overall amplitude of variability is roughly the same in the
short- and long-wavelength IUE bands, but
the ultraviolet spectral shape does vary significantly during the campaign.
Comparing Figures~3 and 7, we can see that the relation between intensity
and spectral shape is not simple. While the source flux doubled in the first
ten days, the spectral index increased briefly
and then returned to roughly its initial
level. In the latter half of the month, while the source intensity remained
high, the spectral index steepened.
During the intensive monitoring period,
the flux-spectral index relationship is
similarly complicated.

Spectral variability is a strong diagnostic of any emission process.
The general sense of the variation in PKS~2155$-$304 is
that the spectrum hardens with increasing intensity,
as expected for nonthermal processes where increases (decreases) result from
acceleration (radiative losses) of the nonthermal electron population.
The apparent temporal lag ($\sim1$~day) between change in spectral shape
and change in intensity (Fig.~9) is more difficult to interpret.

The specific accretion disk model proposed by Wandel~\& Urry~(1991),
which was constructed to fit the ultraviolet spectral shape and
estimated ultraviolet variability time scales of PKS~2155$-$304,
is ruled out by
the close correlation between optical and ultraviolet IUE light curves,
since in that model a large fraction of the optical flux had to be produced
independently of the ultraviolet flux.
Additional arguments against the disk model for PKS~2155$-$304 are based on the
observed wavelength-dependent optical polarization
(increasing to the blue; Smith~\& Sitko~1991; Allen~\etal1992)
and the extremely rapid changes in polarization (Smith~\etal1992).

In contrast,
the ultraviolet variability is consistent with what is expected
from a relativistic jet. Specifically,
Celotti, Maraschi~\& Treves~(1991) consider the variability
characteristics of a range of jet models, assuming that an increase in
emissivity is triggered by a signal traveling along the jet, without
affecting the local particle spectra. In this case
it is possible to produce
ultraviolet variability with very small spectral changes.
All the models predict much stronger X-ray variability; both the
relative amplitude and the lag, if any, are strong constraints
on the details of the model.
No detailed calculations were attempted here because the
associated X-ray variability is the strongest constraint;
for further discussion, see Edelson~\etal(1993).

The variability could also be caused by gravitational micro-lensing.
Ostriker and Vietri~(1985, 1989) have argued that the redshift distribution
of BL~Lacertae objects can be explained by incorrect identification
of the lensing galaxies as the host galaxies of the BL~Lacs, with
the true background objects being OVV quasars.
Specific candidates for micro-lensing have been suggested for some
of the radio-selected BL~Lac objects on the basis of their
variability characteristics (Stickel~\etal1991).
If the size of the emitting region is independent of wavelength,
the variability will be achromatic, as is approximately the case
with the IUE data for PKS~2155$-$304. Rapid variations ($t_{var}\ll1$~year)
through micro-lensing are possible but require the relative
source/lens velocities to be extremely high, such as the
superluminal motion that results from an aligned relativistic jet.
Calculations of the amplification pattern due to micro-lensing that
include the macro-lensing shear can even produce a quasi-periodic signal
(Wambsganss, Paczynski~\& Katz~1990). Minor spectral changes could be
explained by intrinsic effects in the jet, while the bulk of the flux
variations could be due to micro-lensing.

\section 5.\ CONCLUSIONS

In November 1991, a large multiwavelength\break campaign was devoted to
monitoring the UV-bright BL~Lac object PKS~2155$-$304, with coverage
from radio through X-ray wavelengths.
Ultraviolet and optical observations were made with the IUE
satellite on a daily basis throughout the month.
Quasi-continuous coverage, with $\sim96$-minute time resolution,
was undertaken during a 4.6-day period in the middle of the month.
The latter data are well-sampled, in the sense that
the fastest flares, with time scales $\sim1$~day,
were clearly resolved.

The month-long observations show a doubling of the flux in about
10~days, with smaller, more rapid flares superimposed on this general trend.
The well-resolved flares during the intensive monitoring
have an apparent quasi-periodicity of $\sim0.7$~days,
although the data train was too short to confirm the reality of the period.
Such flares might be expected from disturbances propagating
along magnetic field lines in a jet (Camenzind~\& Krockenberger~1992).

The large amplitude and short time scale of the fastest
observed variability probably requires relativistic beaming
(depending on whether emission at other wavelengths varies comparably,
so that the bolometric correction of 10 is applicable).
In the absence of relativistic effects, however,
a simple estimate identifying the periodicity with a Keplerian
orbit at the smallest stable orbit around a central black hole
limits the black hole mass to $\aplt1.3 \times 10^8~M_\odot$.

The short- and long-wavelength ultraviolet light curves are well
correlated with each other, and with the optical
light curve deduced from the Fine Error Sensor (FES) on IUE.
The formal lag is zero but the cross-correlation is
asymmetric in the sense that shorter wavelength emission leads the longer,
suggesting that the loss time scales
are faster for the shorter wavelength emission,
as expected in the synchrotron process.

Small but significant spectral variability is detected.
The ultraviolet spectral index and intensity are inversely correlated
(harder spectra corresponding to higher intensity),
but the change in index leads the change in intensity by one or
two days.
The inverse correlation between spectral index and intensity
is consistent with the nonthermal acceleration processes expected
in relativistic plasmas, so that the present results are
again consistent with relativistic jet models.

The accretion disk scenario of Wandel~\& Urry (1991) is ruled out
by the similarity of the optical and ultraviolet light curves.
The general character of the variability is consistent
with the expectations of relativistic jet models
(Celotti~\etal1991; Camenzind~\& Krockenberger~1992), and
we cannot rule out the importance of gravitational micro-lensing.

\acknow

We are grateful to the IUE project for important contributions to the
observations, particularly Bruce~McCollum (Goddard) and Richard~Monier
(Vilspa), who were in charge of scheduling the observations;
the Goddard Telescope Operators (Jim~Cap-\break linger, Andy~Groebner,
Charlie~Loomis, Scott~Snell, Tom~Walker, and Daryl~Weinstein)
and Resident Astronomers (Richard~Arquilla, Michael~Carini,
Martin~England,
Cathy~Mansperger, Jeff~Newmark, Mar-\break io~Perez, and Lloyd~Rawley);
and the Vilspa staff (Domitilla~de~Martino, John~Fernley,
Rosario~Gonzalez,
and Antonio~Talavera).
The IUE Regional Data Analysis Facility was instrumental in making prompt
access to the data possible, and Lyla~Taylor helpfully provided a
replacement tape when one of our original data tapes was unreadable.
The Vilspa staff were also extremely helpful in providing a second
copy of the Vilspa images on tape.
We thank M.~Begelman, M.~Donahue, A.~Lawrence,
F.~Makino, D.~Maoz, E.~Rosenblatt, A.~Sadun,
P.~Smith, \hbox{W.-H.}~Sun, E.~Tanzi, and H.~C.~Thomas
for their support of this project, and the IUE Peer Review Panel for
its generous allocation of time.
CMU acknowledges Rick~Shafer and Jerry~Kriss for helpful
discussions about error analysis, and Max~Camenzind and Gopal~Krishna for
helpful discussions about the interpretation.
We thank Jay~Lockman for providing H~I measurements toward PKS~2155$-$304
in advance of publication.
This work was supported in part by NASA grant NAG~5-1034.
A.~V.~F.\ acknowledges support from NSF grant AST-8957063 and
NASA grant NAG~5-1800.

\refs

Allen, R.~G., Smith, P.~S., Angel, J.~R.~P., Miller, B.~W., Anderson,
S.~F., \& Margon, B.\ 1992, ApJ, in press

Barylak, M., Wasatonic, R., \& Imhoff, C.\ 1984, ESA IUE Newsletter,
20, 201

Bevington, P.~R.\ 1969, Data Reduction and Error Analysis for the
Physical Sciences, (New York: McGraw-Hill), p.~169

Blandford, R., \& Rees, M.~J.\ 1978, in Pittsburgh Conference on
BL~Lac
Objects, ed.\ A.~M.~Wolfe, (U.~of Pittsburgh), p.~328

Bohlin, R.\ 1988, NASA IUE Newsletter, 35, 141

Bohlin, R., \& Grillmair, C.~J.\ 1988, ApJS, 68, 487

Bohlin, R., Harris, A.~W., Holm, A.~V., \& Gry, C.\ 1990, ApJS, 73, 413

Bowyer, S., Brodie, J., Clarke, J.~T., \& Henry, J.~P.\
1984, ApJ, 278, L103

Brinkmann, W., \etal1992, in preparation

Camenzind, M., \& Krockenberger, M.\ 1992, A\&A, 255, 59

Canizares, C.~R., \& Kruper, J.~S.\ 1984, ApJ, 278, L99

Carini, M., \& Miller, H.~R.\ 1992, ApJ, 385, 482

Carrasco, L., Dultzin-Hacyan, D., \& Cruz-Gonzalez, I.\ 1985,
Nature, 314, 146

Cassatella, A., Lloyd, C., \& Riestra, G.\ 1988,
NASA IUE Newsletter, 35, 225

Celotti, A., Maraschi, L., \& Treves, A.\ 1991, ApJ, 377, 403

Clavel, J., \etal1991, ApJ, 366, 64

Courvoisier, T., \etal1992, in preparation

Crenshaw, M.~D., Bruegman, O.~W., \& Norman, D.~J.\ 1990, PASP, 102, 463

Edelson, R.~A.\ 1992, ApJ, 401, 516

Edelson, R.~A., \etal1991, ApJ, 372, L9

Edelson, R.~A., \etal1993, in preparation

Edelson, R.~A., \& Krolik, J.~H.\ 1988, ApJ, 333, 646

Edelson, R., Pike, G.~F., Saken, J.~M., Kinney, A., \& Shull,
J.~M.\ 1992, ApJS, 83, 1

Elvis, M., Lockman, F.~J., \& Wilkes, B.~J.\ 1989, AJ, 97, 777

Fabian, A.~C.\ 1979, Proc.\ Roy.\ Soc., 366, 449

Falomo, R., Giraud, E., Maraschi, L., Melnick, J., Tanzi, E.~G.,
\& Treves, A.\ 1991, ApJ, 380, L67

Feigelson, E., \etal1986, ApJ, 302, 337

Finley, D., \etal1993, in preparation

George, I.~M., Warwick, R.~S., \& Bromage, G.~E.\ 1988, MNRAS, 232, 793

Ghisellini, G., Maraschi, L., \& Treves, A.\ 1985, A\&A, 146, 204

Holm, A.~V., \& Crabb, W.~G.\ 1979, NASA IUE Newsletter, 7, 40

Hutter, D.~J., \& Mufson, S.~L.\ 1986, ApJ, 301, 50

Kinney, A.~L., Bohlin, R.~C., \& Neill, J.~D.\ 1991, PASP, 103, 694

K\"onigl, A.\ 1981, ApJ, 243, 700

Lockman, F.~J., \& Savage, B.~D.\ 1993, in preparation

Lockman, F.~J., Jahoda, K., \& McCammon, D.\ 1986, ApJ, 302, 432

Madejski, G.\ 1985, Ph.D.\ Thesis, Harvard University

Malina, R.~F., \& Bowyer, C.~S.\ 1992, EUVE Newsletter, Vol.~2,
No.~6

Maraschi, L., Tagliaferri, G., Tanzi, E.~G., \& Treves, A.\
1986, ApJ, 304, 637

Marscher, A.~P.\ 1980, ApJ, 235, 386

Morini, M., \etal1986, ApJ, 306, L71

Mufson, S.~L., Hutter, D.~J., Kondo, Y., Urry, C.~M., \&
Wisniewski, W.~Z.\ 1990, ApJ, 354, 116

Ostriker, J.~P., \& Vietri, M.\ 1985, Nature, 318, 446

Ostriker, J.~P., \& Vietri, M.\ 1990, Nature, 344, 45

Padovani, P., \& Urry, C.~M.\ 1990, ApJ, 356, 75

Perez, M., \& Loomis, C.\ 1991, Record of the Meeting of the International
Ultraviolet Explorer User's Committee, (CSC/TM-91/6142), p.~I-3

Pounds, K.~A., Allen, D.~J., Barber, C., Barstow, M.~A.\ 1992,
MNRAS, in press

Press, W.\ 1978, Comm.\ Ap., 7, 103

Quirrenbach, A., Witzel, A., Krichbaum, T.~P., Hummel, A., Alberdi, A.,
\& Schalinski, C.\ 1989, Nature, 337, 442

Quirrenbach, A., Witzel, A., Wagner, S., Sanchez-Pons, F.,
Krichbaum, T.~P., Wegner, R., Anton, K., Erkens, U., Haehnelt, M.,
Zensus, J.~A., \& Johnston, K.~J.\ 1991, ApJ, 372, L71

Schneider, P., \& Weiss, A.\ 1987, A\&A, 171, 49

Seaton, M.~J.\ 1979, MNRAS, 187, 73p

Sembay, S., Warwick, R.~S., Urry, C.~M., Sokoloski, J., George,
I.~M., Makino, F., \& Ohashi, T.\ 1992, ApJ, submitted

Shull, J.~M., \& Van~Steenberg, M.\ 1985, ApJ, 294, 599

Smith, P.~S., \& Sitko, M.~L.\ 1991, ApJ, 383, 580

Smith, P.~S., Hall, P.~B., Allen, R.~G., \& Sitko, M.~L.\ 1992,
ApJ, in press

Snyder, W.~A., \etal1980, ApJ, L11

Stark, A.~A., Gammie, C.~F., Wilson, R.~W., Bally, J., Linke, R.~A.,
Heiles, C., \& Hurwitz, M.\ 1992, ApJS, 79, 77

Stickel, M., Fried, J.~W., \& K\"uhr, H.\ 1988, A\&A, 191, L16

Stickel, M., Padovani, P., Urry, C.~M., Fried, J.~W., \& K\"uhr,
H.\ 1991, ApJ, 374, 431

Teays, T., \& Garhart, M.\ 1990, NASA IUE Newsletter, 41, 94

Treves, A., \etal1989, ApJ, 341, 733

Ulrich, M.-H., Hackney, K.~R.~H., Hackney, R.~L., \& Kondo, Y.\
1984, ApJ, 276, 466

Urry, C.~M., Kondo, Y., Hackney, K.~R.~H., \& Hackney, R.~L.\ 1988,
ApJ, 330, 791

Urry, C.~M., \& Mushotzky, R.~F.\ 1982, ApJ, 253, 38

Urry, C.~M., \& Reichert, G.~A.\ 1988, IUE Newsletter, 34, 95

Wagner, S., \& Witzel, A.\ 1992, in Extragalactic Radio Sources,
eds.\
J.~Roland, H.~Sol, G.~Pelletier, (Cambridge: Cambridge Univ. Press),
p.~59

Walter, R., \& Courvoisier, T.~J.-L.\ 1991, A\&A, 250, 312

Wambsganss, J., Paczynski, B., \& Katz, N.\ 1990, ApJ, 352, 407

Wandel, A., \& Urry, C.~M.\ 1991, ApJ, 367, 78

Worrall, D.~M., \etal1986, ApJ, 303, 589

\endrefs

\figcaps

\fig Fig.\ 1.---Representative spectra from among the 201 spectra
in the data set. In both cases the source intensities are approximately the
median values for the month-long monitoring campaign, and the exposure times
are typical.
Regions excluded from the power-law fitting are shown as dotted lines.
Best-fit power law models, weighted according to the SWET error vector
plotted in the panel at bottom, are shown as smooth curves; the spectra have
not been de-reddened, so the fits shown
differ from those presented in Table~2.
{\it (a)} An SWP spectrum obtained on 14~November~1991, with
$\chi_\nu^2 = 3.44$ for the best-fit power-law model.
{\it (b)} An LWP spectrum obtained on 10~November~1991, with
$\chi_\nu^2 = 2.25$ for the best-fit power law.

\fig Fig.\ 2.---Cumulative distribution of normalized reduced
chi-squared values for power-law fits to the 99 LWP
({\it dark histogram}) and 102 SWP spectra
({\it light histogram}),
compared to the expected probability $P(\chi_\nu^2)$ ({\it curve}).
The mean of the distribution is 1, by definition, since the SWET error
vector was corrected by the square-root of the means of the original
distributions ($\langle \chi_\nu^2 \rangle = 3.72$ for the SWP and
$\langle \chi_\nu^2 \rangle = 2.42$ for the LWP).
In both cases, a K-S test gives
a reasonable probability ($P_{K-S}>0.5$) that the observed and expected
distributions are the same.

\fig Fig.\ 3.---Ultraviolet light curves of PKS~2155$-$304.
{\it (a)}~The full month-long
light curve, with fitted LWP fluxes at 2800~\AA\ ({\it open
squares}) and SWP fluxes at 1400~\AA\ ({\it filled circles}) on the same scale.
Both long- and short-wavelength fluxes doubled during the month, with
no apparent lag. The four SWP fluxes that are uncertain due to anomalously
short exposure times are circled.
{\it (b)}~Expanded view of the intensive monitoring period, during which
IUE observations were nearly continuous. The LWP scale is at left, the
SWP scale at right. Many rapid flares
have clearly been well-sampled. Five cycles with period $\sim0.7$~day
can be seen, but the reality of the
quasi-periodicity cannot be established without a longer data train.

\fig Fig.\ 4.---Optical light curves deduced from the IUE FES monitor
for {\it (a)}~the full month and {\it (b)}~the intensive monitoring period.
FES counts were converted to approximate V magnitude according to the
prescription of Perez~\& Loomis~(1991), which takes into account the
contribution of scattered light. Typical error bars of
$\sim0.08$~mag (see \S~2.5) are shown in each panel.

\fig Fig.\ 5.---Discrete cross-correlations between SWP flux
and longer-wavelength flux for the intensive monitoring period.
({\it filled circles}) SWP flux at 1400~\AA\ versus LWP flux at
2800~\AA;
({\it open circles}) SWP versus FES flux.
The light curves are highly correlated, with peak at zero lag
(with an upper limit of less than a few hours), and at the same time
asymmetric (the excess positive correlation at positive lags corresponds
to the short-wavelength emission leading longer-wavelength emission).

\fig Fig.\ 6.---Discrete auto-correlation functions for SWP flux
at 1400~\AA\ ({\it filled circles}), LWP flux at 2800~\AA\
({\it open squares}), and FES flux ({\it open circles}).
{\it (a)}~Calculated for the full 30-day data set, which is probably
undersampled. There is correlation out to several ($\sim5$) days.
The amplitude of the FES autocorrelation is low because of the larger
relative errors in the photometry.
{\it (b)}~Calculated for the 4.6-day intensive monitoring period only.
The SWP and LWP autocorrelation functions are very similar, and both show
a broad peak at a lag of about 0.7 days, and a first harmonic at about 1.5
days.

\fig Fig.\ 7.---Variability of the combined spectral index
{\it (a)}~over the full month and {\it (b)}~during the
intensive monitoring period.
The model of a constant spectral shape has
$\chi_\nu^2 =4.7$, so is strongly rejected.

\fig Fig.\ 8.---Ultraviolet flux versus spectral index.
{\it (a)}~$F_{2000}$ versus $\alpha_C$ for the full month {\it (open circles)}
and for the intensive monitoring period {\it (filled circles)}.
{\it (b)}~$\Delta F_{2000}$ versus $\Delta \alpha_C$ for
the full month {\it (open circles)}
and for the intensive monitoring period {\it (filled circles)}.
There are no significant correlations according to a Spearman Rank-Order test.

\fig Fig.\ 9.---Discrete cross-correlation of spectral index versus intensity
for the intensive monitoring period. ({\it filled circles}) $F_{1400}$
vs.\ $\alpha_{SWP}$; ({\it open squares}) $F_{2800}$ vs.\ $\alpha_{LWP}$.
Note the dip at $-1$~day. The negative lag means
change in the spectral index precedes change in intensity, and the
negative correlation amplitude means that spectral hardening (decreasing
$\alpha$) is associated with increasing intensity.

\fig Fig.\ 10.---Histograms of spectral index distributions for SWP
{\it (solid line)} and LWP {\it (dot-dash line)}.  {\it (a)}~Fits
presented in Table~2, corrected for reddening using $E(B-V)=0.034$~mag;
{\it (b)}~raw data, not corrected for reddening;
{\it (c)}~fits after correction with $E(B-V)=0.05$~mag.
Given the reddening uncertainty, no spectral curvature can be claimed.
The reddening correction adopted (panel {\it a)}
brings the SWP and LWP distributions most closely into agreement.

\endfigcaps

\pageno=15

\centerline{\bf TABLE 1. LOG OF IUE OBSERVATIONS OF PKS
2155$-$304}
\trule
\tabskip=2em plus 1em minus 1em
\halign to
\hsize{\hfil#\hfil&\hfil#\hbox{\hskip28pt}&\hfil#\hfil&\hfil#\hfil&\hfil#\hfil
\cr
&\multispan1\hfil Start Time\hfil&Exposure&V Magnitude&Observatory\cr
IUE&\multispan1\hfil (UT)\hfil&Time&(from FES&(Goddard or\cr
Image&\multispan1\hfil(Day of Nov
91)\hfil&(min)&counts)&Vilspa)\cr
\mrule
LWP21607&       1.832&         30&        13.29&        G\cr
SWP42969&       1.860&         55&        13.29&        G\cr
LWP21608&       1.905&         30&        13.29&        G\cr
SWP42970&       1.930&         55&        13.34&        G\cr
LWP21609&       1.971&         25&        13.24&        G\cr
SWP42971&       1.993&         55&        13.33&        G\cr
LWP21610&       2.036&         25&        13.26&        G\cr
SWP42972&       2.059&         55&        13.25&        G\cr
LWP21611&       2.101&         25&        13.29&        G\cr
LWP21616&       2.825&         25&        13.18&        G\cr
SWP42979&       2.847&         55&        13.15&        G\cr
LWP21617&       2.889&         25&        13.18&        G\cr
SWP42980&       2.913&         55&        13.29&        G\cr
LWP21625&       3.834&         25&        13.18&        G\cr
SWP42995&       3.857&         55&        13.15&        G\cr
LWP21626&       3.905&         25&        13.18&        G\cr
SWP42996&       3.929&         35&        13.20&        G\cr
LWP21636&       4.817&         25&        12.99&        G\cr
SWP43008&       4.842&         55&        12.98&        G\cr
LWP21637&       4.888&         25&        12.99&        G\cr
SWP43009&       4.913&         55&        13.02&        G\cr
LWP21644&       5.828&         25&        13.03&        G\cr
SWP43017&       5.853&         55&        12.97&        G\cr
LWP21645&       5.903&         25&        12.95&        G\cr
SWP43018&       5.928&         55&        12.94&        G\cr
LWP21652&       6.813&         25&         --- &        G\cr
SWP43025&       6.985&         55&        13.05&        G\cr
LWP21653&       7.031&         25&        13.05&        G\cr
SWP43026&       7.056&         55&        13.08&        G\cr
LWP21654&       7.103&         25&        13.08&        G\cr
LWP21667&       9.041&         25&        12.94&        G\cr
SWP43040&       9.177&         55&        12.96&        G\cr
LWP21668&       9.224&         25&        12.89&        G\cr
SWP43041&       9.252&         55&        12.92&        G\cr
LWP21673&       9.675&         25&        12.86&        V\cr
SWP43047&       9.700&         55&        12.85&        V\cr
LWP21674&       9.742&         25&        12.89&        V\cr
\phantom{$^a$}SWP 43048$^a$&      9.765&         25&        12.85&        V\cr
LWP21683&      10.663&         25&        12.83&        V\cr
SWP43054&      10.685&         55&        12.81&        V\cr
LWP21684&      10.728&         25&        12.87&        V\cr
SWP43055&      10.751&         55&        12.78&        G\cr
LWP21685&      10.797&         25&        12.76&        G\cr
SWP43056&      10.821&         51&        12.84&        G\cr
LWP21686&      10.864&         25&        12.81&        G\cr
SWP43057&      10.886&         55&        12.80&        G\cr
LWP21687&      10.929&         25&        12.76&        G\cr
\brule
\noalign{\vfill\eject
\centerline{\bf TABLE 1. LOG OF IUE OBSERVATIONS OF PKS
2155$-$304 (cont'd)}
\trule}
&\multispan1\hfil Start Time\hfil&Exposure&V Magnitude&Observatory\cr
IUE&\multispan1\hfil (UT)\hfil&Time&(from FES&(Goddard or\cr
Image&\multispan1\hfil(Day of Nov
91)\hfil&(min)&counts)&Vilspa)\cr
\mrule
SWP43058&      10.953&         52&        12.80&        G\cr
LWP21688&      10.996&         25&        12.79&        G\cr
SWP43059&      11.018&         55&        12.77&        G\cr
LWP21689&      11.062&         25&        12.75&        G\cr
SWP43060&      11.085&         52&        12.77&        G\cr
LWP21690&      11.128&         25&        12.75&        G\cr
SWP43061&      11.152&         51&        12.76&        G\cr
LWP21691&      11.195&         23&        12.72&        G\cr
SWP43062&      11.219&         51&        12.74&        G\cr
LWP21692&      11.260&         25&        12.70&        G\cr
SWP43063&      11.285&         55&        12.72&        G\cr
LWP21693&      11.328&         25&        12.70&        G\cr
SWP43064&      11.353&         50&        12.71&        G\cr
SWP43065&      11.815&         57&        12.73&        G\cr
LWP21696&      11.862&         25&         --- &        G\cr
SWP43066&      11.889&         45&        12.69&        G\cr
LWP21697&      11.924&         27&        12.69&        G\cr
SWP43067&      11.947&         58&        12.66&        G\cr
LWP21698&      11.993&         25&         --- &        G\cr
SWP43068&      12.014&         57&        12.68&        G\cr
LWP21699&      12.058&         27&        12.68&        G\cr
SWP43069&      12.081&         57&        12.68&        G\cr
LWP21700&      12.125&         27&        12.66&        G\cr
SWP43070&      12.150&         54&        12.72&        G\cr
LWP21701&      12.192&         25&        12.66&        G\cr
SWP43071&      12.215&         55&        12.70&        G\cr
LWP21702&      12.258&         28&        12.72&        G\cr
\phantom{$^a$}SWP 43073$^a$&      12.510&         13&        12.74&        V\cr
LWP21704&      12.525&         25&        12.84&        V\cr
SWP43074&      12.549&         55&        12.73&        V\cr
LWP21705&      12.592&         25&        12.65&        V\cr
SWP43075&      12.615&         55&        12.70&        V\cr
LWP21706&      12.658&         25&        12.73&        V\cr
SWP43076&      12.686&         49&        12.73&        V\cr
LWP21707&      12.725&         25&        12.69&        V\cr
SWP43077&      12.748&         55&        12.63&        G\cr
LWP21708&      12.791&         25&        12.68&        G\cr
SWP43078&      12.814&         55&        12.68&        G\cr
LWP21709&      12.858&         25&        12.69&        G\cr
SWP43079&      12.880&         55&        12.69&        G\cr
LWP21710&      12.922&         28&        12.72&        G\cr
SWP43080&      12.946&         57&        12.69&        G\cr
LWP21711&      12.991&         25&        12.73&        G\cr
SWP43081&      13.015&         53&        12.78&        G\cr
LWP21712&      13.061&         25&        12.73&        G\cr
SWP43082&      13.083&         51&        12.75&        G\cr
LWP21713&      13.123&         25&        12.74&        G\cr
\brule
\noalign{\vfill\eject
\centerline{\bf TABLE 1. LOG OF IUE OBSERVATIONS OF PKS
2155$-$304 (cont'd)}
\trule}
&\multispan1\hfil Start Time\hfil&Exposure&V Magnitude&Observatory\cr
IUE&\multispan1\hfil (UT)\hfil&Time&(from FES&(Goddard or\cr
Image&\multispan1\hfil(Day of Nov
91)\hfil&(min)&counts)&Vilspa)\cr
\mrule
SWP43083&      13.150&         51&        12.75&        G\cr
LWP21714&      13.189&         25&        12.72&        G\cr
SWP43084&      13.212&         55&        12.75&        G\cr
LWP21715&      13.258&         25&        12.72&        G\cr
SWP43085&      13.280&         55&        12.75&        G\cr
LWP21716&      13.322&         25&        12.78&        G\cr
SWP43086&      13.343&         65&        12.81&        G\cr
\phantom{$^a$}SWP 43088$^a$&   13.498&         27&        12.73&        V\cr
LWP21717&      13.523&         25&        12.75&        V\cr
SWP43089&      13.554&         42&        12.76&        V\cr
LWP21718&      13.590&         25&        12.80&        V\cr
SWP43090&      13.612&         55&        12.71&        V\cr
LWP21719&      13.656&         25&        12.82&        V\cr
SWP43091&      13.679&         55&        12.84&        V\cr
LWP21720&      13.723&         25&        12.83&        V\cr
SWP43092&      13.748&         52&        12.84&        G\cr
LWP21721&      13.788&         26&        12.80&        G\cr
SWP43093&      13.811&         56&        12.78&        G\cr
LWP21722&      13.854&         27&        12.86&        G\cr
SWP43094&      13.877&         57&        12.88&        G\cr
LWP21723&      13.921&         27&        12.87&        G\cr
SWP43095&      13.948&         52&        12.91&        G\cr
LWP21724&      13.988&         27&        12.87&        G\cr
SWP43096&      14.011&         57&        12.89&        G\cr
LWP21725&      14.054&         27&        12.91&        G\cr
SWP43097&      14.077&         57&        12.91&        G\cr
LWP21726&      14.121&         27&        12.89&        G\cr
SWP43098&      14.144&         57&        12.87&        G\cr
LWP21727&      14.189&         25&        12.64&        G\cr
SWP43099&      14.213&         53&        12.86&        G\cr
LWP21728&      14.255&         30&        12.80&        G\cr
LWP21730&      14.521&         25&        12.87&        V\cr
SWP43101&      14.544&         55&        12.84&        V\cr
LWP21731&      14.587&         25&        12.79&        V\cr
SWP43102&      14.610&         55&        12.77&        V\cr
LWP21732&      14.654&         25&        12.83&        V\cr
SWP43103&      14.681&         49&        12.77&        V\cr
LWP21733&      14.721&         25&        12.71&        V\cr
SWP43104&      14.744&         55&        12.79&        G\cr
LWP21734&      14.786&         27&        12.69&        G\cr
SWP43105&      14.810&         55&        12.74&        G\cr
LWP21735&      14.852&         27&        12.72&        G\cr
SWP43106&      14.875&         57&        12.75&        G\cr
LWP21736&      14.919&         27&        12.70&        G\cr
SWP43107&      14.942&         57&        12.73&        G\cr
LWP21737&      14.985&         27&        12.74&        G\cr
SWP43108&      15.009&         57&        12.75&        G\cr
\brule
\noalign{\vfill\eject
\centerline{\bf TABLE 1. LOG OF IUE OBSERVATIONS OF PKS
2155$-$304 (cont'd)}
\trule}
&\multispan1\hfil Start Time\hfil&Exposure&V Magnitude&Observatory\cr
IUE&\multispan1\hfil (UT)\hfil&Time&(from FES&(Goddard or\cr
Image&\multispan1\hfil(Day of Nov
91)\hfil&(min)&counts)&Vilspa)\cr
\mrule
LWP21738&      15.051&         27&        12.74&        G\cr
SWP43109&      15.074&         58&        12.71&        G\cr
LWP21739&      15.119&         27&        12.72&        G\cr
SWP43110&      15.144&         54&        12.77&        G\cr
LWP21740&      15.185&         27&        12.78&        G\cr
SWP43111&      15.207&         57&        12.73&        G\cr
LWP21741&      15.250&         35&        12.72&        G\cr
LWP21744&      15.816&         25&        12.77&        G\cr
LWP21747&      16.168&         25&        12.71&        G\cr
SWP43114&      16.190&         55&        12.67&        G\cr
LWP21748&      16.232&         25&        12.64&        G\cr
SWP43115&      16.254&         40&        12.64&        G\cr
LWP21755&      16.821&         25&        12.66&        G\cr
SWP43121&      16.841&         55&        12.66&        G\cr
LWP21756&      16.882&         25&        12.61&        G\cr
SWP43122&      16.903&         65&        12.64&        G\cr
LWP21768&      17.819&         25&        12.73&        G\cr
SWP43135&      17.841&         55&        12.68&        G\cr
LWP21769&      17.883&         25&        12.67&        G\cr
SWP43136&      17.908&         60&        12.68&        G\cr
SWP43145&      18.815&         55&        12.74&        G\cr
LWP21777&      18.859&         27&        12.76&        G\cr
SWP43146&      18.884&         55&        12.74&        G\cr
LWP21778&      18.928&         33&        12.73&        G\cr
LWP21786&      20.014&         25&        12.73&        G\cr
SWP43157&      20.035&         55&        12.78&        G\cr
LWP21787&      20.078&         27&        12.81&        G\cr
SWP43158&      20.103&         33&        12.79&        G\cr
SWP43164&      20.695&         55&        12.69&        V\cr
LWP21793&      20.740&         25&        12.69&        V\cr
\phantom{$^a$}SWP 43165$^a$&   20.766&         25&        12.69&        V\cr
SWP43174&      21.673&         55&        12.69&        V\cr
LWP21799&      21.718&         25&        12.74&        V\cr
SWP43175&      21.747&         55&        12.65&        V\cr
SWP43184&      22.662&         55&        12.60&        V\cr
LWP21810&      22.707&         25&         --- &        V\cr
SWP43185&      22.730&         47&        12.76&        V\cr
LWP21811&      22.767&         25&         --- &        V\cr
SWP43192&      23.683&         55&        12.59&        V\cr
LWP21828&      23.729&         25&        12.68&        V\cr
SWP43193&      23.753&         44&        12.60&        V\cr
\phantom{$^b$}SWP 43211$^b$&   24.684&         55&        12.78&        V\cr
\phantom{$^b$}LWP 21837$^b$&   24.750&         32&        12.73&        V\cr
SWP43220&      25.679&         55&        12.83&        V\cr
LWP21847&      25.723&         25&        12.77&        V\cr
SWP43221&      25.749&         50&        12.79&        V\cr
SWP43230&      26.678&         55&        12.58&        V\cr\brule
\noalign{\vfill\eject
\centerline{\bf TABLE 1. LOG OF IUE OBSERVATIONS OF PKS
2155$-$304 (cont'd)}
\trule}
&\multispan1\hfil Start Time\hfil&Exposure&V Magnitude&Observatory\cr
IUE&\multispan1\hfil (UT)\hfil&Time&(from FES&(Goddard or\cr
Image&\multispan1\hfil(Day of Nov
91)\hfil&(min)&counts)&Vilspa)\cr
\mrule
LWP21856&      26.724&         25&        12.54&        V\cr
SWP43231&      26.751&         45&        12.56&        V\cr
SWP43236&      27.677&         55&        12.59&        V\cr
LWP21864&      27.725&         25&        12.56&        V\cr
SWP43237&      27.747&         50&        12.54&        V\cr
LWP21877&      28.832&         25&        11.82&        G\cr
SWP43246&      28.858&         45&        12.59&        G\cr
LWP21878&      28.895&         25&        12.61&        G\cr
SWP43247&      28.919&         45&        12.62&        G\cr
SWP43260&      29.824&         55&        12.55&        G\cr
LWP21888&      29.867&         25&        12.58&        G\cr
SWP43261&      29.893&         50&        12.61&        G\cr
LWP21889&      29.932&         25&        12.57&        G\cr
\brule
}
\vskip12pt

\noindent {\it Notes }--- $^a$Unusually short exposure time;
$^b$Interrupted exposure

\vfill\eject

\centerline{\bf TABLE 2. POWER-LAW FITS TO DEREDDENED SPECTRA}

\trule

\tabskip=2em plus1em minus1em

\halign to
\hsize{\hfil#\hfil&\hfil#\hskip18pt&\hfil#\hfil&\hfil#\hfil&\hfil#\hfil&\hfil#\hfil&
\hfil#\hfil\cr
&\multispan1\hfil Observation\hfil&$F_{1400}$&$\sigma_{F_{1400}}$\cr
IUE&\multispan1\hfil Midpoint (UT)\hfil&\multispan2\hrulefill
&\multispan2\hfil Spectral Index\hfil&\cr
Image&\multispan1\hfil (Day of Nov 91)\hfil&\multispan2\hfil
($\times10^{-14}$ ergs cm$^{-2}$ s$^{-1}$ \AA$^{-1}$)\hfil&$\alpha_{SWP}$
&$\sigma_{\alpha_{SWP}}$&$\chi_\nu^2$
\cr
\mrule
SWP 42969&   1.87912&   11.6800&   0.1616&   0.881&   0.057&   3.917\cr
SWP 42970&   1.94888&   11.8200&   0.1630&   0.855&   0.057&   3.398\cr
SWP 42971&   2.01254&   12.0900&   0.1663&   0.777&   0.057&   3.707\cr
SWP 42972&   2.07773&   12.0600&   0.1657&   0.961&   0.057&   3.713\cr
SWP 42979&   2.86603&   13.7800&   0.1880&   0.860&   0.053&   2.992\cr
SWP 42980&   2.93207&   14.2300&   0.1938&   0.740&   0.053&   4.751\cr
SWP 42995&   3.87643&   14.0900&   0.1918&   0.760&   0.053&   3.718\cr
SWP 42996&   3.94092&   14.0000&   0.1975&   0.879&   0.064&   3.213\cr
SWP 43008&   4.86099&   15.8900&   0.2152&   0.771&   0.049&   3.569\cr
SWP 43009&   4.93195&   15.9400&   0.2153&   0.768&   0.049&   3.775\cr
SWP 43017&   5.87164&   16.2400&   0.2190&   0.847&   0.049&   3.525\cr
SWP 43018&   5.94727&   16.5800&   0.2232&   0.736&   0.049&   3.839\cr
SWP 43025&   7.00403&   14.9100&   0.2016&   0.789&   0.049&   5.203\cr
SWP 43026&   7.07547&   14.7000&   0.1991&   0.885&   0.049&   3.949\cr
SWP 43040&   9.19641&   18.0200&   0.2402&   0.734&   0.045&   3.845\cr
SWP 43041&   9.27063&   18.5000&   0.2476&   0.710&   0.045&   3.327\cr
SWP 43047&   9.71866&   18.9900&   0.2537&   0.772&   0.045&   3.749\cr
SWP 43048&   9.77417&   18.7100&   0.2633&   0.887&   0.060&   3.834\cr
SWP 43054&  10.70386&   20.7100&   0.2749&   0.688&   0.041&   4.008\cr
SWP 43055&  10.77008&   20.8600&   0.2774&   0.677&   0.045&   4.105\cr
SWP 43056&  10.83871&   20.8500&   0.2782&   0.759&   0.045&   3.364\cr
SWP 43057&  10.90527&   21.1300&   0.2810&   0.750&   0.045&   3.783\cr
SWP 43058&  10.97137&   20.9500&   0.2790&   0.743&   0.045&   3.591\cr
SWP 43059&  11.03726&   21.1200&   0.2804&   0.722&   0.041&   4.201\cr
SWP 43060&  11.10294&   21.2100&   0.2840&   0.758&   0.045&   3.133\cr
SWP 43061&  11.16980&   21.7500&   0.2890&   0.711&   0.041&   3.692\cr
SWP 43062&  11.23703&   22.2600&   0.2988&   0.753&   0.049&   3.439\cr
SWP 43063&  11.30371&   22.1100&   0.3033&   0.678&   0.057&   4.099\cr
SWP 43064&  11.37048&   22.6500&   0.3023&   0.750&   0.045&   4.225\cr
SWP 43065&  11.83447&   22.0900&   0.2919&   0.727&   0.041&   4.024\cr
SWP 43066&  11.90454&   22.3400&   0.2974&   0.732&   0.045&   3.853\cr
SWP 43067&  11.96689&   22.3000&   0.2944&   0.779&   0.041&   2.887\cr
SWP 43068&  12.03357&   23.0000&   0.3041&   0.722&   0.041&   4.573\cr
SWP 43069&  12.10080&   22.3500&   0.2957&   0.772&   0.041&   4.018\cr
SWP 43070&  12.16840&   21.7800&   0.2892&   0.821&   0.045&   3.384\cr
SWP 43071&  12.23428&   21.1200&   0.2804&   0.809&   0.045&   3.665\cr
SWP 43073&  12.51471&   18.8900&   0.2947&   1.016&   0.090&   2.946\cr
SWP 43074&  12.56787&   20.6700&   0.2749&   0.760&   0.045&   3.883\cr
SWP 43075&  12.63367&   20.7800&   0.2763&   0.776&   0.045&   4.197\cr
SWP 43076&  12.70255&   21.0700&   0.2808&   0.784&   0.045&   3.841\cr
SWP 43077&  12.76688&   21.3900&   0.2794&   0.780&   0.038&   3.616\cr
SWP 43078&  12.83356&   21.3100&   0.2828&   0.797&   0.041&   3.679\cr
SWP 43079&  12.89929&   21.8800&   0.2903&   0.797&   0.041&   3.138\cr
SWP 43080&  12.96588&   21.8600&   0.2896&   0.819&   0.041&   3.899\cr
SWP 43081&  13.03345&   21.4900&   0.2857&   0.860&   0.045&   4.173\cr
SWP 43082&  13.10092&   20.9100&   0.2788&   0.813&   0.045&   3.697\cr
SWP 43083&  13.16806&   20.4200&   0.2727&   0.844&   0.045&   3.693\cr
\brule
\noalign{\vfill\eject
\centerline{\bf TABLE 2. POWER-LAW FITS TO DEREDDENED SPECTRA
(cont'd)}
\trule}
&\multispan1\hfil Observation\hfil&$F_{1400}$&$\sigma_{F_{1400}}$\cr
IUE&\multispan1\hfil Midpoint (UT)\hfil&\multispan2\hrulefill
&\multispan2\hfil Spectral Index\hfil&\cr
Image&\multispan1\hfil (Day of Nov 91)\hfil&\multispan2\hfil
($\times10^{-14}$ ergs cm$^{-2}$ s$^{-1}$ \AA$^{-1}$)\hfil&$\alpha_{SWP}$
&$\sigma_{\alpha_{SWP}}$&$\chi_\nu^2$
\cr
\mrule
SWP 43084&  13.23126&   19.9700&   0.2667&   0.764&   0.045&   4.003\cr
SWP 43085&  13.29907&   19.6300&   0.2641&   0.838&   0.049&   3.666\cr
SWP 43086&  13.36594&   19.6400&   0.2568&   0.758&   0.038&   4.387\cr
SWP 43088&  13.50720&   19.1800&   0.2687&   0.942&   0.060&   3.498\cr
SWP 43089&  13.56894&   20.0200&   0.2708&   0.767&   0.049&   3.651\cr
SWP 43090&  13.63153&   20.3000&   0.2708&   0.718&   0.045&   4.344\cr
SWP 43091&  13.69809&   20.0100&   0.2668&   0.833&   0.045&   4.346\cr
SWP 43092&  13.76599&   19.6500&   0.2625&   0.809&   0.045&   3.698\cr
SWP 43093&  13.83084&   19.3000&   0.2574&   0.755&   0.045&   3.078\cr
SWP 43094&  13.89709&   18.9600&   0.2529&   0.740&   0.045&   3.208\cr
SWP 43095&  13.96570&   18.3800&   0.2467&   0.739&   0.049&   3.861\cr
SWP 43096&  14.03030&   18.4000&   0.2462&   0.755&   0.045&   3.680\cr
SWP 43097&  14.09653&   18.0600&   0.2418&   0.779&   0.045&   3.569\cr
SWP 43098&  14.16336&   17.9200&   0.2399&   0.798&   0.045&   3.755\cr
SWP 43099&  14.23135&   17.7800&   0.2387&   0.833&   0.049&   3.510\cr
SWP 43101&  14.56296&   18.7000&   0.2497&   0.770&   0.045&   4.200\cr
SWP 43102&  14.62903&   19.0700&   0.2543&   0.802&   0.045&   4.055\cr
SWP 43103&  14.69815&   19.5200&   0.2610&   0.846&   0.045&   3.618\cr
SWP 43104&  14.76279&   19.9900&   0.2659&   0.854&   0.045&   3.592\cr
SWP 43105&  14.82904&   20.4300&   0.2717&   0.819&   0.045&   3.717\cr
SWP 43106&  14.89505&   20.6500&   0.2740&   0.766&   0.041&   3.698\cr
SWP 43107&  14.96176&   20.2500&   0.2690&   0.816&   0.045&   3.427\cr
SWP 43108&  15.02841&   20.4200&   0.2711&   0.729&   0.041&   3.590\cr
SWP 43109&  15.09451&   20.1700&   0.2678&   0.733&   0.041&   3.994\cr
SWP 43110&  15.16229&   19.5300&   0.2603&   0.786&   0.045&   3.821\cr
SWP 43111&  15.22717&   19.7900&   0.2633&   0.807&   0.045&   4.813\cr
SWP 43114&  16.20871&   21.4200&   0.2837&   0.813&   0.041&   3.963\cr
SWP 43115&  16.26831&   21.4600&   0.2895&   0.832&   0.049&   4.100\cr
SWP 43121&  16.86020&   22.1400&   0.2929&   0.727&   0.041&   3.614\cr
SWP 43122&  16.92563&   22.8200&   0.3005&   0.668&   0.041&   3.813\cr
SWP 43135&  17.86005&   21.3700&   0.2844&   0.778&   0.045&   3.734\cr
SWP 43136&  17.92896&   21.3600&   0.2828&   0.739&   0.041&   3.600\cr
SWP 43145&  18.83432&   19.3500&   0.2573&   0.847&   0.045&   3.267\cr
SWP 43146&  18.90332&   18.7700&   0.2503&   0.920&   0.045&   4.066\cr
SWP 43157&  20.05450&   18.6300&   0.2483&   0.848&   0.045&   3.303\cr
SWP 43158&  20.11401&   18.0000&   0.2478&   0.944&   0.057&   2.800\cr
SWP 43164&  20.71451&   20.1300&   0.2676&   0.867&   0.045&   3.645\cr
SWP 43165&  20.77444&   19.4600&   0.2728&   0.955&   0.060&   4.011\cr
SWP 43174&  21.69171&   18.5600&   0.2478&   0.939&   0.045&   3.911\cr
SWP 43175&  21.76627&   18.7400&   0.2501&   0.932&   0.045&   3.583\cr
SWP 43184&  22.68073&   19.4800&   0.2598&   0.811&   0.045&   3.747\cr
SWP 43185&  22.74625&   19.4000&   0.2606&   0.826&   0.049&   3.755\cr
SWP 43192&  23.70206&   21.3400&   0.2834&   0.816&   0.041&   4.284\cr
SWP 43193&  23.76813&   21.0300&   0.2825&   0.897&   0.049&   3.511\cr
SWP 43211&  24.70337&   20.0400&   0.2668&   0.782&   0.045&   3.758\cr
SWP 43220&  25.69824&   19.6400&   0.2622&   0.819&   0.045&   3.393\cr
SWP 43221&  25.76685&   19.6500&   0.2636&   0.801&   0.045&   3.576\cr
\brule
\noalign{\vfill\eject
\centerline{\bf TABLE 2. POWER-LAW FITS TO DEREDDENED SPECTRA
(cont'd)}
\trule}
&\multispan1\hfil Observation\hfil&$F_{2800}$&$\sigma_{F_{2800}}$\cr
IUE&\multispan1\hfil Midpoint (UT)\hfil&\multispan2\hrulefill
&\multispan2\hfil Spectral Index\hfil&\cr
Image&\multispan1\hfil (Day of Nov 91)\hfil&\multispan2\hfil
($\times10^{-14}$ ergs cm$^{-2}$ s$^{-1}$ \AA$^{-1}$)\hfil&$\alpha_{LWP}$
&$\sigma_{\alpha_{LWP}}$&$\chi_\nu^2$
\cr
\mrule
SWP 43230&  26.69736&   22.8400&   0.3027&   0.791&   0.041&   3.955\cr
SWP 43231&  26.76675&   23.4000&   0.3128&   0.842&   0.045&   3.383\cr
SWP 43236&  27.69647&   20.9400&   0.2781&   0.806&   0.041&   3.942\cr
SWP 43237&  27.76450&   20.8300&   0.2780&   0.886&   0.045&   3.161\cr
SWP 43246&  28.87381&   21.9100&   0.2923&   0.909&   0.045&   4.232\cr
SWP 43247&  28.93491&   22.1400&   0.2949&   0.871&   0.045&   3.695\cr
SWP 43260&  29.84335&   22.9300&   0.3041&   0.727&   0.041&   4.265\cr
SWP 43261&  29.90997&   22.7000&   0.3019&   0.832&   0.045&   4.116\cr
LWP 21607&   1.84244&    4.6230&   0.0669&   0.697&   0.142&   1.775\cr
LWP 21608&   1.91510&    4.7150&   0.0682&   0.934&   0.142&   2.376\cr
LWP 21609&   1.97968&    4.7130&   0.0706&   0.972&   0.161&   2.204\cr
LWP 21610&   2.04449&    4.7990&   0.0716&   0.938&   0.161&   2.152\cr
LWP 21611&   2.10934&    4.8180&   0.0713&   0.740&   0.154&   1.949\cr
LWP 21616&   2.83383&    5.2770&   0.0766&   0.585&   0.146&   2.255\cr
LWP 21617&   2.89771&    5.3020&   0.0774&   0.927&   0.150&   2.065\cr
LWP 21625&   3.84235&    5.2430&   0.0762&   0.490&   0.146&   2.397\cr
LWP 21626&   3.91318&    5.4070&   0.0783&   0.692&   0.142&   2.564\cr
LWP 21636&   4.82599&    6.1480&   0.0872&   0.985&   0.131&   2.039\cr
LWP 21637&   4.89694&    6.1810&   0.0880&   0.964&   0.135&   2.373\cr
LWP 21644&   5.83676&    6.2380&   0.0882&   0.755&   0.127&   2.357\cr
LWP 21645&   5.91180&    6.3510&   0.0897&   0.735&   0.127&   2.052\cr
LWP 21652&   6.82202&    6.1910&   0.0876&   0.778&   0.131&   2.385\cr
LWP 21653&   7.03946&    5.8710&   0.0833&   0.678&   0.131&   2.117\cr
LWP 21654&   7.11139&    5.9320&   0.0848&   0.879&   0.135&   2.019\cr
LWP 21667&   9.04929&    6.9190&   0.0968&   0.847&   0.124&   2.400\cr
LWP 21668&   9.23254&    7.1040&   0.0996&   1.155&   0.124&   2.538\cr
LWP 21673&   9.68405&    7.3060&   0.1019&   0.854&   0.120&   2.375\cr
LWP 21674&   9.75076&    7.3250&   0.1022&   0.887&   0.120&   2.521\cr
LWP 21683&  10.67126&    7.8460&   0.1086&   0.903&   0.116&   2.805\cr
LWP 21684&  10.73682&    7.8530&   0.1083&   0.751&   0.112&   1.748\cr
LWP 21685&  10.80591&    7.9180&   0.1095&   0.854&   0.116&   2.285\cr
LWP 21686&  10.87241&    7.9760&   0.1102&   0.908&   0.112&   2.306\cr
LWP 21687&  10.93814&    7.8680&   0.1088&   0.653&   0.116&   1.980\cr
LWP 21688&  11.00476&    7.9780&   0.1101&   0.820&   0.112&   2.488\cr
LWP 21689&  11.07086&    8.0130&   0.1105&   0.770&   0.112&   2.558\cr
LWP 21690&  11.13641&    8.1230&   0.1117&   0.723&   0.112&   2.376\cr
LWP 21691&  11.20306&    8.3940&   0.1164&   0.901&   0.116&   2.414\cr
LWP 21692&  11.26898&    8.5300&   0.1217&   0.748&   0.135&   2.290\cr
LWP 21693&  11.33670&    8.5630&   0.1206&   0.693&   0.127&   2.503\cr
LWP 21696&  11.87045&    8.5250&   0.1169&   0.707&   0.109&   2.732\cr
LWP 21697&  11.93353&    8.5700&   0.1167&   0.641&   0.105&   1.939\cr
LWP 21698&  12.00150&    8.7160&   0.1193&   0.772&   0.109&   2.496\cr
LWP 21699&  12.06717&    8.7480&   0.1187&   0.663&   0.101&   3.072\cr
LWP 21700&  12.13443&    8.6080&   0.1171&   0.795&   0.105&   2.691\cr
LWP 21701&  12.20102&    8.5050&   0.1170&   0.968&   0.109&   2.687\cr
LWP 21702&  12.26740&    8.2260&   0.1127&   0.771&   0.109&   2.509\cr
LWP 21704&  12.53397&    8.1640&   0.1121&   0.655&   0.109&   3.045\cr
\brule
\noalign{\vfill\eject
\centerline{\bf TABLE 2. POWER-LAW FITS TO DEREDDENED SPECTRA
(cont'd)}
\trule}
&\multispan1\hfil Observation\hfil&$F_{2800}$&$\sigma_{F_{2800}}$\cr
IUE&\multispan1\hfil Midpoint (UT)\hfil&\multispan2\hrulefill
&\multispan2\hfil Spectral Index\hfil&\cr
Image&\multispan1\hfil (Day of Nov 91)\hfil&\multispan2\hfil
($\times10^{-14}$ ergs cm$^{-2}$ s$^{-1}$ \AA$^{-1}$)\hfil&$\alpha_{LWP}$
&$\sigma_{\alpha_{LWP}}$&$\chi_\nu^2$
\cr
\mrule
LWP 21705&  12.60086&    8.2070&   0.1128&   0.875&   0.109&   2.260\cr
LWP 21706&  12.66708&    8.1920&   0.1128&   0.752&   0.112&   2.374\cr
LWP 21707&  12.73361&    8.2920&   0.1142&   0.842&   0.112&   2.662\cr
LWP 21708&  12.79990&    8.6660&   0.1193&   1.241&   0.112&   4.104\cr
LWP 21709&  12.86646&    8.3230&   0.1144&   0.786&   0.109&   2.425\cr
LWP 21710&  12.93216&    8.5110&   0.1156&   0.837&   0.101&   2.322\cr
LWP 21711&  13.00006&    8.4760&   0.1165&   0.957&   0.109&   2.386\cr
LWP 21712&  13.06924&    8.2690&   0.1138&   0.934&   0.112&   2.498\cr
LWP 21713&  13.13214&    8.0630&   0.1109&   0.776&   0.112&   2.394\cr
LWP 21714&  13.19809&    7.9180&   0.1094&   0.897&   0.112&   2.441\cr
LWP 21715&  13.26642&    7.6570&   0.1071&   0.678&   0.124&   2.438\cr
LWP 21716&  13.33105&    7.5940&   0.1061&   0.792&   0.120&   2.559\cr
LWP 21717&  13.53159&    7.7690&   0.1075&   0.882&   0.116&   2.648\cr
LWP 21718&  13.59872&    7.8320&   0.1080&   0.687&   0.112&   2.260\cr
LWP 21719&  13.66470&    7.9080&   0.1094&   0.862&   0.116&   2.380\cr
LWP 21720&  13.73138&    7.6380&   0.1059&   0.768&   0.116&   2.534\cr
LWP 21721&  13.79739&    7.6450&   0.1057&   0.892&   0.112&   2.570\cr
LWP 21722&  13.86356&    7.3160&   0.1009&   0.685&   0.112&   2.374\cr
LWP 21723&  13.93060&    7.2740&   0.1007&   0.896&   0.116&   2.952\cr
LWP 21724&  13.99689&    7.1100&   0.0983&   0.713&   0.112&   2.379\cr
LWP 21725&  14.06299&    7.1500&   0.0990&   0.931&   0.116&   2.976\cr
LWP 21726&  14.12988&    7.0930&   0.0985&   0.921&   0.116&   2.143\cr
LWP 21727&  14.19751&    7.0190&   0.0979&   0.784&   0.120&   2.156\cr
LWP 21728&  14.26501&    7.0270&   0.0965&   0.700&   0.109&   2.421\cr
LWP 21730&  14.52951&    7.2790&   0.1015&   0.889&   0.120&   2.396\cr
LWP 21731&  14.59549&    7.5110&   0.1039&   0.780&   0.116&   2.830\cr
LWP 21732&  14.66269&    7.6730&   0.1063&   0.896&   0.116&   2.427\cr
LWP 21733&  14.72964&    7.8020&   0.1080&   0.830&   0.116&   2.123\cr
LWP 21734&  14.79572&    8.0160&   0.1096&   0.905&   0.109&   2.764\cr
LWP 21735&  14.86139&    8.0320&   0.1098&   0.761&   0.109&   2.442\cr
LWP 21736&  14.92819&    7.9950&   0.1096&   0.914&   0.109&   2.460\cr
LWP 21737&  14.99469&    8.0120&   0.1099&   0.935&   0.109&   2.581\cr
LWP 21738&  15.06076&    7.9780&   0.1091&   0.804&   0.109&   2.580\cr
LWP 21739&  15.12793&    7.8050&   0.1070&   0.774&   0.109&   2.395\cr
LWP 21740&  15.19437&    7.7950&   0.1072&   0.859&   0.112&   2.462\cr
LWP 21741&  15.26242&    8.0120&   0.1083&   0.999&   0.097&   2.718\cr
LWP 21744&  15.82483&    7.6290&   0.1056&   0.883&   0.116&   2.398\cr
LWP 21747&  16.17667&    8.3520&   0.1147&   0.914&   0.109&   2.562\cr
LWP 21748&  16.24106&    8.4620&   0.1161&   0.816&   0.109&   2.403\cr
LWP 21755&  16.82968&    8.4450&   0.1160&   0.768&   0.109&   2.654\cr
LWP 21756&  16.89108&    8.7600&   0.1199&   0.958&   0.109&   2.106\cr
LWP 21768&  17.82776&    8.1080&   0.1118&   0.803&   0.112&   2.280\cr
LWP 21769&  17.89197&    8.3060&   0.1142&   0.766&   0.112&   1.918\cr
LWP 21777&  18.86792&    7.8310&   0.1076&   0.889&   0.112&   2.308\cr
LWP 21778&  18.93918&    7.8110&   0.1009&   0.719&   0.064&   2.642\cr
LWP 21786&  20.02225&    7.4050&   0.1031&   1.067&   0.120&   2.139\cr
LWP 21787&  20.08722&    7.4130&   0.1020&   0.834&   0.112&   2.238\cr
\brule
\noalign{\vfill\eject
\centerline{\bf TABLE 2. POWER-LAW FITS TO DEREDDENED SPECTRA
(cont'd)}
\trule}
&\multispan1\hfil Observation\hfil&$F_{2800}$&$\sigma_{F_{2800}}$\cr
IUE&\multispan1\hfil Midpoint (UT)\hfil&\multispan2\hrulefill
&\multispan2\hfil Spectral Index\hfil&\cr
Image&\multispan1\hfil (Day of Nov 91)\hfil&\multispan2\hfil
($\times10^{-14}$ ergs cm$^{-2}$ s$^{-1}$ \AA$^{-1}$)\hfil&$\alpha_{LWP}$
&$\sigma_{\alpha_{LWP}}$&$\chi_\nu^2$
\cr
\mrule
LWP 21793&  20.74850&    8.3290&   0.1141&   0.845&   0.109&   2.544\cr
LWP 21799&  21.72665&    7.6930&   0.1064&   0.947&   0.116&   1.758\cr
LWP 21810&  22.71548&    7.9510&   0.1094&   0.909&   0.112&   2.110\cr
LWP 21811&  22.77560&    7.8670&   0.1086&   0.837&   0.116&   2.269\cr
LWP 21828&  23.73749&    8.7470&   0.1198&   0.914&   0.109&   2.281\cr
LWP 21837&  24.76144&    8.0990&   0.1099&   0.934&   0.101&   2.703\cr
LWP 21847&  25.73145&    7.8290&   0.1092&   0.882&   0.120&   2.639\cr
LWP 21856&  26.73236&    9.3260&   0.1270&   0.841&   0.105&   2.797\cr
LWP 21864&  27.73346&    8.4550&   0.1157&   0.938&   0.109&   2.219\cr
LWP 21877&  28.84067&    9.0620&   0.1236&   0.916&   0.105&   2.538\cr
LWP 21878&  28.90332&    9.1780&   0.1248&   0.920&   0.105&   2.252\cr
LWP 21888&  29.87552&    9.2210&   0.1256&   0.668&   0.105&   2.619\cr
LWP 21889&  29.94113&    9.1820&   0.1250&   0.855&   0.105&   2.427\cr
\brule
\noalign{\vfill\eject}
}

\centerline{\bf TABLE 3.\ POWER-LAW FITS TO MERGED SWP-LWP SPECTRA}

\trule

\tabskip=2em plus1em minus1em

\halign to \hsize{\hfil#\hfil&\hfil#\hfil&\hfil#\hskip23pt&\hfil#&\hfil#\hfil&
\hfil#\hfil&\hfil#\hfil&\hfil#\hfil\cr
\multispan2\hfil Spectral Pair\hfil&\multispan1\hfil Observation\hfil&
$F_{2000}$&$\sigma_{F_{2000}}$&&&\cr
\multispan2\hfil Image Numbers\hfil&\multispan1\hfil Midpoint
(UT)\hfil&\multispan2
\hrulefill&\multispan2\hfil Spectral Index\hfil&\cr
SWP&LWP&\multispan1\hfil (Day of Nov 91)\hfil&\multispan2\hfil
($\times 10^{-14}$ ergs cm$^{-2}$ s$^{-1}$ \AA$^{-1}$)\hfil
&$\alpha_C$&$\sigma_{\alpha_C}$&$\chi^2_\nu$\cr
\mrule
42969&    21607&    1.86078&      5.9859&      0.0779&    0.6240&
0.0234&    0.833\cr
42970&    21608&    1.93201&      6.0642&      0.0788&    0.6490&
0.0234&    0.881\cr
42971&    21609&    1.99609&      6.1090&      0.0796&    0.6200&
0.0239&    0.819\cr
42972&    21610&    2.06110&      6.2548&      0.0816&    0.6260&
0.0243&    1.002\cr
42972&    21611&    2.09354&      6.2716&      0.0817&    0.6290&
0.0237&    0.936\cr
42979&    21616&    2.84991&      6.8937&      0.0896&    0.5780&
0.0227&    0.831\cr
42980&    21617&    2.91489&      6.9639&      0.0902&    0.5520&
0.0220&    1.035\cr
42995&    21625&    3.85938&      6.8988&      0.0894&    0.5390&
0.0218&    0.908\cr
42996&    21626&    3.92706&      7.0487&      0.0918&    0.5830&
0.0243&    0.923\cr
43008&    21636&    4.84351&      7.9526&      0.1027&    0.6130&
0.0214&    0.880\cr
43009&    21637&    4.91443&      8.0223&      0.1035&    0.6110&
0.0207&    0.951\cr
43017&    21644&    5.85419&      8.0999&      0.1044&    0.5800&
0.0209&    1.901\cr
43018&    21645&    5.92953&      8.2564&      0.1066&    0.5950&
0.0214&    0.874\cr
43025&    21652&    6.91302&      7.8411&      0.1012&    0.7200&
0.0214&    1.070\cr
43026&    21653&    7.05746&      7.3442&      0.0949&    0.6300&
0.0218&    3.953\cr
43026&    21654&    7.09344&      7.6212&      0.0984&    0.6620&
0.0209&    0.898\cr
43040&    21667&    9.12286&      8.9753&      0.1153&    0.6020&
0.0191&    0.960\cr
43041&    21668&    9.25159&      9.1723&      0.1179&    0.6140&
0.0200&    0.988\cr
43047&    21673&    9.70135&      9.4772&      0.1219&    0.6010&
0.0198&    0.926\cr
43048&    21674&    9.76245&      9.4906&      0.1233&    0.6070&
0.0234&    1.011\cr
43054&    21683&   10.68756&     10.1939&      0.1307&    0.5880&
0.0188&    1.089\cr
43055&    21684&   10.75345&     10.2047&      0.1309&    0.5790&
0.0188&    0.826\cr
43056&    21685&   10.82233&     10.3228&      0.1324&    0.5810&
0.0186&    0.899\cr
43057&    21686&   10.88885&     10.4067&      0.1333&    0.5740&
0.0184&    0.975\cr
43058&    21687&   10.95477&     10.2889&      0.1320&    0.5610&
0.0188&    0.855\cr
43059&    21688&   11.02100&     10.3994&      0.1331&    0.5760&
0.0184&    0.961\cr
43060&    21689&   11.08691&     10.4654&      0.1343&    0.5690&
0.0191&    0.903\cr
43061&    21690&   11.15311&     10.6500&      0.1365&    0.5540&
0.0186&    0.944\cr
43062&    21691&   11.22003&     10.9667&      0.1409&    0.5690&
0.0195&    0.967\cr
43063&    21692&   11.28635&     10.9578&      0.1420&    0.6220&
0.0227&    0.904\cr
43064&    21693&   11.35358&     11.1509&      0.1431&    0.5750&
0.0195&    0.953\cr
43065&    21696&   11.85248&     11.0276&      0.1411&    0.6090&
0.0181&    1.014\cr
43066&    21697&   11.91904&     11.1152&      0.1423&    0.5960&
0.0181&    0.849\cr
43067&    21698&   11.98419&     11.2635&      0.1441&    0.6220&
0.0177&    0.860\cr
43068&    21699&   12.05035&     11.3657&      0.1452&    0.5850&
0.0175&    1.171\cr
43069&    21700&   12.11761&     11.1887&      0.1430&    0.5980&
0.0177&    1.057\cr
43070&    21701&   12.18469&     11.0182&      0.1411&    0.6190&
0.0184&    1.024\cr
43071&    21702&   12.25085&     10.6448&      0.1362&    0.6130&
0.0179&    0.958\cr
43073&    21704&   12.52435&     10.3198&      0.1367&    0.7240&
0.0294&    0.928\cr
43074&    21705&   12.58435&     10.5344&      0.1350&    0.6540&
0.0186&    0.889\cr
43075&    21706&   12.65039&     10.5557&      0.1353&    0.6360&
0.0186&    0.955\cr
43076&    21707&   12.71811&     10.6899&      0.1370&    0.6350&
0.0188&    0.996\cr
43077&    21708&   12.78345&     11.0509&      0.1414&    0.6890&
0.0181&    1.334\cr
43078&    21709&   12.85004&     10.7665&      0.1378&    0.6190&
0.0181&    0.959\cr
43079&    21710&   12.91574&     11.0285&      0.1410&    0.6100&
0.0177&    0.903\cr
43080&    21711&   12.98297&     10.9998&      0.1406&    0.6080&
0.0177&    1.027\cr
43081&    21712&   13.05136&     10.7693&      0.1378&    0.5900&
0.0181&    1.056\cr
\brule
}
\vfill\eject

\centerline{\bf TABLE 3.\ POWER-LAW FITS TO MERGED SWP-LWP
SPECTRA (cont'd)}

\trule

\tabskip=2em plus1em minus1em

\halign to \hsize{\hfil#\hfil&\hfil#\hfil&\hfil#\hskip20pt&\hfil#&\hfil#\hfil&
\hfil#\hfil&\hfil#\hfil&\hfil#\hfil\cr
\multispan2\hfil Spectral Pair\hfil&\multispan1\hfil Observation\hfil&
$F_{2000}$&$\sigma_{F_{2000}}$&&&\cr
\multispan2\hfil Image Number\hfil&\multispan1\hfil Midpoint
(UT)\hfil&\multispan2
\hrulefill&\multispan2\hfil Spectral Index\hfil&\cr
SWP&LWP&\multispan1\hfil (Day of Nov 91)\hfil&\multispan2\hfil
($\times 10^{-14}$ ergs cm$^{-2}$ s$^{-1}$ \AA$^{-1}$)\hfil
&$\alpha_C$&$\sigma_{\alpha_C}$&$\chi^2_\nu$\cr
\mrule
43082&    21713&   13.11655&     10.4873&      0.1344&    0.5960&
0.0186&    0.963\cr
43083&    21714&   13.18307&     10.2873&      0.1320&    0.6040&
0.0191&    0.981\cr
43084&    21715&   13.24884&      9.9549&      0.1281&    0.5920&
0.0195&    0.982\cr
43085&    21715&   13.28275&      9.9238&      0.1276&    0.6100&
0.0200&    0.950\cr
43086&    21716&   13.34851&      9.8422&      0.1261&    0.6090&
0.0186&    1.051\cr
43088&    21717&   13.51941&      9.9996&      0.1296&    0.6500&
0.0227&    1.000\cr
43089&    21718&   13.58383&     10.1054&      0.1300&    0.6240&
0.0200&    0.883\cr
43090&    21719&   13.64813&     10.1821&      0.1307&    0.6300&
0.0193&    1.011\cr
43091&    21720&   13.71472&     10.0029&      0.1284&    0.5730&
0.0193&    1.140\cr
43092&    21721&   13.78168&      9.8835&      0.1269&    0.6160&
0.0191&    0.948\cr
43093&    21722&   13.84720&      9.5388&      0.1223&    0.5750&
0.0188&    0.864\cr
43094&    21723&   13.91385&      9.4330&      0.1211&    0.6020&
0.0191&    1.000\cr
43095&    21724&   13.98129&      9.1971&      0.1182&    0.6110&
0.0195&    0.940\cr
43096&    21725&   14.04663&      9.2319&      0.1185&    0.6220&
0.0193&    1.051\cr
43097&    21726&   14.11322&      9.1581&      0.1175&    0.6320&
0.0193&    0.882\cr
43098&    21727&   14.18042&      9.0881&      0.1167&    0.6220&
0.0193&    0.917\cr
43099&    21728&   14.24817&      9.0826&      0.1166&    0.6270&
0.0193&    0.943\cr
43101&    21730&   14.54623&      9.4185&      0.1210&    0.6190&
0.0198&    1.007\cr
43102&    21731&   14.61224&      9.6806&      0.1241&    0.6330&
0.0186&    1.001\cr
43103&    21732&   14.68042&      9.9503&      0.1278&    0.6210&
0.0193&    0.977\cr
43104&    21733&   14.74622&     10.1550&      0.1301&    0.6070&
0.0186&    0.970\cr
43105&    21734&   14.81238&     10.3669&      0.1327&    0.6250&
0.0181&    1.074\cr
43106&    21735&   14.87823&     10.3875&      0.1329&    0.6170&
0.0181&    0.939\cr
43107&    21736&   14.94498&     10.3210&      0.1321&    0.6350&
0.0181&    0.961\cr
43108&    21737&   15.01154&     10.2987&      0.1318&    0.6410&
0.0181&    0.946\cr
43109&    21738&   15.07764&     10.2264&      0.1310&    0.6520&
0.0184&    0.968\cr
43110&    21739&   15.14511&     10.0141&      0.1283&    0.6570&
0.0186&    0.923\cr
43111&    21740&   15.21075&     10.0667&      0.1289&    0.6320&
0.0184&    1.118\cr
43111&    21741&   15.24481&     10.2396&      0.1309&    0.6800&
0.0177&    1.182\cr
43114&    21747&   16.19269&     10.8351&      0.1385&    0.6150&
0.0179&    1.084\cr
43115&    21748&   16.25470&     10.9505&      0.1405&    0.6240&
0.0191&    1.024\cr
43121&    21755&   16.84494&     10.9845&      0.1406&    0.5900&
0.0181&    0.973\cr
43122&    21756&   16.90836&     11.3194&      0.1445&    0.6140&
0.0175&    0.917\cr
43135&    21768&   17.84390&     10.5614&      0.1355&    0.5770&
0.0188&    0.913\cr
43136&    21769&   17.91046&     10.7159&      0.1372&    0.6230&
0.0181&    0.799\cr
43145&    21777&   18.85114&     10.0257&      0.1285&    0.6720&
0.0184&    0.906\cr
43146&    21778&   18.92126&      9.9314&      0.1270&    0.7020&
0.0177&    1.028\cr
43157&    21786&   20.03839&      9.5462&      0.1227&    0.6460&
0.0193&    0.900\cr
43158&    21787&   20.10062&      9.4811&      0.1220&    0.6800&
0.0209&    0.820\cr
43164&    21793&   20.73151&     10.6023&      0.1359&    0.7040&
0.0186&    0.940\cr
43165&    21793&   20.76147&     10.5371&      0.1364&    0.7350&
0.0225&    0.957\cr
43174&    21799&   21.70917&      9.8364&      0.1263&    0.6960&
0.0193&    0.892\cr
43175&    21799&   21.74646&      9.8784&      0.1268&    0.6800&
0.0188&    0.919\cr
43184&    21810&   22.69812&     10.1398&      0.1300&    0.6910&
0.0188&    0.901\cr
43185&    21811&   22.76093&     10.0724&      0.1295&    0.6740&
0.0198&    0.930\cr
43192&    21828&   23.71979&     11.1318&      0.1424&    0.6990&
0.0181&    0.981\cr
43193&    21828&   23.75281&     11.1296&      0.1428&    0.7070&
0.0191&    0.908\cr
\brule
}
\vfill\eject

\centerline{\bf TABLE 3.\ POWER-LAW FITS TO MERGED SWP-LWP
SPECTRA (cont'd)}

\trule

\tabskip=2em plus1em minus1em

\halign to \hsize{\hfil#\hfil&\hfil#\hfil&\hfil#\hskip20pt&\hfil#&\hfil#\hfil&
\hfil#\hfil&\hfil#\hfil&\hfil#\hfil\cr
\multispan2\hfil Spectral Pair\hfil&\multispan1\hfil Observation\hfil&
$F_{2000}$&$\sigma_{F_{2000}}$&&&\cr
\multispan2\hfil Image Number\hfil&\multispan1\hfil Midpoint
(UT)\hfil&\multispan2
\hrulefill&\multispan2\hfil Spectral Index\hfil&\cr
SWP&LWP&\multispan1\hfil (Day of Nov 91)\hfil&\multispan2\hfil
($\times 10^{-14}$ ergs cm$^{-2}$ s$^{-1}$ \AA$^{-1}$)\hfil
&$\alpha_C$&$\sigma_{\alpha_C}$&$\chi^2_\nu$\cr
\mrule
43211&    21837&   24.73242&     10.3553&      0.1326&    0.6780&
0.0181&    1.011\cr
43220&    21847&   25.71484&     10.0799&      0.1294&    0.6490&
0.0193&    0.981\cr
43221&    21847&   25.74915&     10.0725&      0.1295&    0.6510&
0.0198&    0.997\cr
43230&    21856&   26.71484&     11.8615&      0.1517&    0.6960&
0.0177&    1.000\cr
43231&    21856&   26.74957&     12.0291&      0.1542&    0.6430&
0.0186&    1.029\cr
43236&    21864&   27.71497&     10.8215&      0.1387&    0.6740&
0.0184&    0.950\cr
43237&    21864&   27.74896&     10.8460&      0.1390&    0.6710&
0.0186&    0.891\cr
43246&    21877&   28.85724&     11.5794&      0.1483&    0.6940&
0.0184&    1.022\cr
43247&    21878&   28.91913&     11.6725&      0.1494&    0.7070&
0.0184&    0.918\cr
43260&    21888&   29.85944&     11.7367&      0.1502&    0.6780&
0.0181&    1.002\cr
43261&    21889&   29.92554&     11.7411&      0.1503&    0.6730&
0.0181&    0.998\cr
\brule
}
\bye